\documentstyle[12pt,psfig]{article}

\newcommand{\lsim}{\raisebox{-0.13cm}{~\shortstack{$<$ \\[-0.07cm] $\sim$}}~}
\newcommand{\gsim}{\raisebox{-0.13cm}{~\shortstack{$>$ \\[-0.07cm] $\sim$}}~}

\newcommand{\ra}{\rightarrow}
\newcommand{\ee}{e^+e^- }
\newcommand{\s}{\\ \vspace*{-3mm} }
\newcommand{\nn}{\noindent}
\newcommand{\non}{\nonumber}
\newcommand{\beq}{\begin{eqnarray}}
\newcommand{\eeq}{\end{eqnarray}}

\newcommand{\tg}{\mbox{tan}}
\newcommand{\tb}{\tan\beta}

\begin{document}

\begin{titlepage}

\begin{flushright}
DESY 95--213\\
KA--TP--10--96\\
IFT-96--09\\
OUTP--96--19P\\
May 1996 \\
\end{flushright}

\vspace{.7cm}

\begin{center}

{\large\sc {\bf  Heavy SUSY Higgs Bosons at $e^+ e^-$ Linear Colliders}}


\vspace{1cm}

{\sc A.~Djouadi$^{1,2}$, J. Kalinowski$^{3}$, P.~Ohmann$^{1,4}$ and
  P.M.~Zerwas$^1$ } 

\vspace{1cm}

$^1$ Deutsches Elektronen--Synchrotron DESY, D-22603 Hamburg, FRG. \\
\vspace{0.3cm}

$^2$ Institut f\"ur Theoretische Physik, Universit\"at Karlsruhe,
D-76128 Karlsruhe, FRG. \\
\vspace{0.3cm}

$^3$ Institute of Theoretical Physics, Warsaw University, PL-00681 Warsaw,
Poland. \\
\vspace{0.3cm}

$^4$ Department of Theoretical Physics, Oxford University, OX1 3NP, 
Oxford, UK. 

\end{center}

\vspace{.6cm}

\begin{abstract}
\normalsize
\noindent

\nn The production mechanisms and decay modes of the heavy neutral and
charged Higgs bosons in the Minimal  Supersymmetric Standard  Model are
investigated at future $e^+ e^-$ colliders in the TeV energy regime. We
generate supersymmetric particle spectra by requiring the MSSM Higgs
potential to produce correct radiative electroweak symmetry breaking,
and we assume a common scalar mass $m_0$, gaugino mass $m_{1/2}$
and trilinear coupling $A$, as well as gauge and Yukawa coupling
unification at the Grand Unification scale. Particular emphasis is put
on the low $\tb$ solution in this scenario where decays of the Higgs
bosons to Standard Model particles compete with decays to supersymmetric
charginos/neutralinos as well as sfermions. In the high $\tb$ case, the
supersymmetric spectrum is either too heavy or the supersymmetric decay
modes are suppressed, since the Higgs bosons decay almost exclusively
into $b$ and $\tau$ pairs. The main production mechanisms for the heavy
Higgs particles are the associated $AH$ production and $H^+H^-$ pair
production with cross sections of the order of a few fb. 
 
\end{abstract}

\end{titlepage}

\setcounter{page}{2}

\setcounter{equation}{0}
\renewcommand{\theequation}{1.\arabic{equation}}

\subsection*{1. Introduction}

Supersymmetric theories \cite{R1,R2} are generally considered to be the
most natural extensions of the Standard Model (SM). This proposition is
based on several points. In these theories, fundamental scalar Higgs
bosons \cite{R3,R4} with low masses can be retained in the context of
high unification scales. Moreover, the prediction \cite{R5} of the
renormalized electroweak mixing angle $\sin^2 \theta_W = 0.2336 \pm
0.0017$, based on the spectrum of the Minimal Supersymmetric Standard
Model (MSSM) \cite{R6}, is in striking agreement with the electroweak
precision data which yield $\sin^2 \theta_W = 0.2314(3)$ \cite{R7}. An
additional attractive feature is provided by the opportunity to generate
the electroweak symmetry breaking radiatively \cite{R8}. If the top
quark mass is in the range between $\sim$ 150 and $\sim$ 200 GeV, the
universal squared Higgs mass parameter at the unification scale
decreases with decreasing energy and becomes negative at the electroweak
scale, thereby breaking the ${\rm SU(2)_L \times U(1)_Y}$ gauge symmetry
while leaving the U(1) electromagnetic and SU(3) color gauge symmetries
intact \cite{R8}. The analysis of the electroweak data prefers a light
Higgs mass \cite{R7,R9} as predicted in supersymmetric theories; however
since the radiative corrections depend only logarithmically on the Higgs
mass \cite{R10}, the dependence is weak and no firm conclusions can yet
be drawn. \s 

The more than doubling the spectrum of states in the MSSM gives rise to
a rather large proliferation of parameters. This number of parameters is
however reduced drastically by embedding the low--energy supersymmetric
theory into a grand unified (GUT) framework. This can be achieved in
supergravity models \cite{R8}, in which the effective low--energy
supersymmetric theory [including the interactions which break
supersymmetry] is described by the following parameters: the common
scalar mass $m_0$, the common gaugino mass $m_{1/2}$, the
trilinear coupling $A$, the bilinear coupling $B$, and the
Higgs--higgsino mass parameter $\mu$. In addition, two parameters are
needed to describe the Higgs sector: one Higgs mass parameter [in
general the mass of the pseudoscalar Higgs boson, $M_A$] and the ratio
of the vacuum expectation values, $\tb =v_2/v_1$, of the two Higgs
doublet fields which break the electroweak symmetry. \s

The number of parameters can be further  reduced by introducing additional
constraints which are based on physically rather natural assumptions: \s

$(i)$
Unification of the $b$ and $\tau$ Yukawa couplings at the GUT scale \cite{R11a}
leads to a correlation between the top quark mass and $\tb$ \cite{R11,R15,R19}.
Adopting the central value of the top mass as measured at the Tevatron
\cite{R12}, $\tb$ is restricted to two narrow ranges around $\tb \sim 1.7$ and
$\tb \sim 50$, with the low $\tb$ solution theoretically somewhat favored 
\cite{R19}. \s

$(ii)$ If the electroweak symmetry is broken radiatively, then the
bilinear coupling $B$ and the Higgs--higgsino mass parameter $\mu$ are
determined up to the sign of $\mu$. [The sign of $\mu$ might be
determined by future precision measurements of the radiative
$b$ decay amplitude.] \s

$(iii)$ It turns out {\it a posteriori} that the physical observables are 
nearly independent of the GUT scale value of the trilinear coupling $A_G$, 
for $|A_G| \lsim 500$ GeV. \s

Mass spectra and couplings of all supersymmetric particles and Higgs
bosons are determined after these steps by just two mass parameters
along with the sign of $\mu$; we shall choose to express our results in
terms of the pseudoscalar Higgs boson $A$ mass $M_A$ and the common GUT
gaugino mass $m_{1/2}$.  \s

In this paper we focus on heavy Higgs particles $A$, $H$ and $H^{\pm}$
with masses of a few hundred GeV, and therefore close to the decoupling
limit \cite{R14}. The pattern of Higgs masses is quite regular in
this limit. While the upper limit on the mass of the lightest CP--even Higgs
boson $h$ is a function of $\tb$ \cite{R14a}, 
\begin{eqnarray}
M_h \lsim 100 \ {\rm to} \ 150 \ {\rm GeV \ \ [ for \ low \ to \ high 
\ \tb}] 
\end{eqnarray}
the heavy Higgs bosons are nearly mass degenerate [c.f. Fig.1]
\beq
M_A & \simeq & M_H  \ \simeq \ M_{H^\pm} 
\eeq
Moreover, the properties of the lightest CP--even Higgs boson $h$ become 
SM--like in this limit. The production of the heavy Higgs bosons becomes 
particularly simple in $e^+ e^-$ collisions; the heavy Higgs bosons can 
only be pair--produced, 
\begin{eqnarray}
e^+ e^- & \rightarrow & A \ H  \\
e^+ e^- & \rightarrow  & H^+ H^- 
\end{eqnarray}
Close to this decoupling limit, the cross section for $H$ Higgs--strahlung $\ee
\ra ZH$ is very small and the cross section for the $WW$ fusion mechanism $\ee
\ra \nu_e \bar{\nu}_e H$ is appreciable only for small values of $\tb$, $\tb
\sim 1$, and relatively small $H$ masses, $M_H \lsim 350$ GeV. The cross
section for $ZZ$ fusion of the $H$ is suppressed by an order of magnitude
compared to $WW$ fusion. The pseudoscalar $A$ particle does not couple to $W/Z$
boson pairs at the tree level. \s 

The decay pattern for heavy Higgs bosons is rather complicated in
general. For large $\tb$ the SM fermion decays prevail. For small $\tb$
this is true above the $t\bar{t}$ threshold of $M_{H,A} \gsim 350$ GeV
for the neutral Higgs bosons and above the $t\bar{b}$ threshold of
$M_{H^\pm} \gsim 180$ GeV for the charged Higgs particles. Below these
mass values many decay channels compete with each other: decays to SM
fermions $f\bar{f}$ [and for $H$ to gauge bosons $VV$], Higgs cascade
decays, chargino/neutralino $\chi_i \chi_j $ decays and decays to 
supersymmetric sfermions $\tilde{f} \tilde{\bar{f}}$
\begin{eqnarray}
H &\rightarrow& f \bar{f}  \ , \ VV \ , \ hh \ , \ \chi_{i} \chi_{j} 
\ , \ \tilde{f} \bar{\tilde{f}} \\
A &\rightarrow& f \bar{f} \ , \ hZ \ , \ \chi_{i} \chi_{j} 
\ , \  \tilde{f} \bar{\tilde{f}} \\ 
H^\pm &\rightarrow& f \bar{f}' \ , \ hW^\pm \ , \ \chi_{i} \chi_{j}
\ , \ \tilde{f} \bar{\tilde{f}} \; ' 
\end{eqnarray}

In this paper, we analyze in detail the decay modes of the heavy Higgs
particles and their production at $\ee$ linear colliders. The analysis will
focus on heavy particles for which machines in the TeV energy range are needed.
The paper is organized in the following way. In the next section we define the
physical set--up of our analysis in the framework of the MSSM embedded into a
minimal supergravity theory. In section 3, we discuss the production cross
sections of the heavy Higgs bosons. In the subsequent sections, we discuss the
widths of the various decay channels and the final Higgs decay products. 
For completeness, analytical expressions for supersymmetric particle masses 
and couplings will be collected in the Appendix.

\setcounter{equation}{0}
\renewcommand{\theequation}{2.\arabic{equation}}

\subsection*{2. The Physical Set--Up}

The Higgs sector of the Minimal Supersymmetric Standard Model is
described at tree-level by the following potential
\begin{eqnarray}
V_0 &=& (m_{H_1}^2+\mu ^2)|H_1|^2+(m_{H_2}^2+\mu
^2)|H_2|^2 - m_3^2(\epsilon_{ij}{H_1}^i{H_2}^j+{\rm h.c.})
\nonumber \\
&& +{1\over 8}(g^2+g^{\prime 2})\left [|H_1|^2-|H_2|^2\right ]^2
+{1\over 2}g^2|H_1^{i*}H_2^i|^2\; 
\end{eqnarray}
The quadratic Higgs terms associated with $\mu$ and the quartic
Higgs terms coming with the electroweak gauge couplings $g$ and $g'$ are
invariant under supersymmetric transformations. $m_{H_1}^{}$,
$m_{H_2}^{}$ and $m_3$ are soft--supersymmetry breaking parameters with
$m_3^2 = B \mu$. $\epsilon_{ij}$ [$i,j=1 , 2$ and $\epsilon_{12}=1$] is
the antisymmetric tensor in two dimensions and $H_1\equiv
(H_1^1,H_1^2)=(H_1^0,H_1^-)$, $H_2\equiv (H_2^1,H_2^2)=(H_2^+,H_2^0)$
are the two Higgs-doublet fields. After the symmetry breaking, three
out of the initially eight degrees of freedom will be absorbed to 
generate the $W^\pm$ and $Z$ masses, leaving a quintet of scalar Higgs
particles: two CP--even Higgs bosons $h$ and $H$, a CP--odd
[pseudoscalar] boson $A$ and two charged Higgs particles $H^\pm$. \s 

Retaining only the [leading] Yukawa couplings of the third generation
\beq
\lambda_t= \frac{\sqrt{2}m_t} {v \sin \beta} \ , \ \ 
\lambda_b= \frac{\sqrt{2}m_b} {v \cos \beta} \ \ {\rm and} \ \ 
\lambda_\tau= \frac{\sqrt{2}m_\tau} {v \cos \beta}
\eeq
where $\tb=v_2/v_1$ [with $v^2=v_1^2+v_2^2$ fixed by the $W$ mass, $v=
246$ GeV] is the ratio of the vacuum expectation values of the fields
$H_2^0$ and $H_1^0$, the superpotential is given in terms of the superfields
$Q=(t,b)$ and $L=(\tau, \nu_\tau)$ by\footnote{Note that 
our convention for the sign of $\mu$ is consistent with Ref.\cite{R2a},
which  is opposite to the one adopted in Ref.\cite{R16}.}
\beq
W= \epsilon_{ij} \left[ \lambda_t H_2^i Q^j t^c + \lambda_b H_1^i Q^j 
b^c + \lambda_\tau H_1^i L^j \tau^c - \mu H^i_1 H_2^j \right]
\eeq
Supersymmetry is broken by introducing the soft--supersymmetry breaking
bino $\tilde{B}$, wino $\tilde{W}^a$ $[a=$1--3] and gluino 
$\tilde{g}^a$ $[a=$1--8] mass 
terms,
\beq
\frac{1}{2} \,M_1 \,\overline{\tilde{B}}\,\tilde{B} \ + \
\frac{1}{2} \,M_2 \,\overline{\tilde{W}}^a \,\tilde{W}^a \ + \
\frac{1}{2} \,M_3 \,\overline{\tilde{g}}^a \,\tilde{g}^a \ , 
\eeq
soft--supersymmetry breaking trilinear couplings,
\beq
\epsilon_{ij} \left[ \lambda_t A_t H_2^i \tilde{Q}^j  \tilde{t}^c + 
\lambda_b A_b H_1^i \tilde{Q}^j \tilde{b}^c + \lambda_\tau A_\tau 
H_1^i \tilde{L}^j  \tilde{\tau}^c - \mu B H^i_1 H_2^j \right]
\eeq
and soft--supersymmetry breaking squark and slepton mass terms
\beq
M_Q^2 [\tilde{t}^*_L\tilde{t}_L +\tilde{b}^*_L\tilde{b}_L] +
M_U^2 \tilde{t}^*_R \tilde{t}_R  + M_D^2 \tilde{b}^*_R \tilde{b}_R+
M_L^2 [\tilde{\tau}^*_L\tilde{\tau}_L +\tilde{\nu_\tau}^*_L\tilde{\nu_\tau}_L] 
+ M_E^2 \tilde{\tau}^*_R \tilde{\tau}_R  \ + \ \cdots
\eeq
where the ellipses stand for the soft mass terms corresponding to the first 
and second generation sfermions. \s

The minimal SUSY--GUT model emerges by requiring at the GUT scale $M_G$: 
\s

\nn $(i)$ the unification of the U(1), SU(2) and SU(3) coupling constants 
$\alpha_i=g_i^2/4\pi$ $[i=1$--3],
\beq
\alpha_3 (M_{\rm G}) = \alpha_2 (M_{\rm G}) = \alpha_1 (M_{\rm G}) =\alpha_G
\eeq
$(ii)$ a common gaugino mass; the $M_i$ with $i=$1--3 at the
electroweak scale are then related through renormalization group equations
(RGEs) to the gauge couplings, 
\beq
M_i = \frac{\alpha_i(M_Z)}{\alpha_G} m_{1/2} \ \longrightarrow 
\ M_3(M_Z)=\frac{\alpha_3(M_Z)} {\alpha_2(M_Z)} M_2(M_Z)
          =\frac{\alpha_3(M_Z)} {\alpha_1(M_Z)} M_1(M_Z)
\eeq
$(iii)$ a universal trilinear coupling $A$
\beq
A_G = A_t (M_{\rm G}) = A_b (M_{\rm G}) = A_\tau (M_{\rm G})
\eeq
$(iii)$ a universal scalar mass $m_0$
\beq
m_0&=& M_Q=M_U=M_D=M_L=M_E \non \\
   &=& m_{H_1}(M_G)=m_{H_2}(M_G) 
\eeq
Besides the three parameters $m_{1/2}, A_G$ and $m_0$ the
supersymmetric sector is described at the GUT scale by the bilinear
coupling $B_G$ and the Higgs--higgsino mass parameter $\mu_G$. The
theoretically attractive assumption that the electroweak symmetry is
broken radiatively constrains the latter two parameters. Indeed,
radiative electroweak symmetry breaking results in two minimization
conditions [see Ref.\cite{R16} for details] of the Higgs
potential; at the low--energy scale in the tree approximation, they are
given by 
\begin{eqnarray}
{1\over 2}M_Z^2&=&{{m_{H_1}^2-m_{H_2}^2\tan ^2\beta }
\over {\tan ^2\beta -1}}-\mu ^2 \;  \\
B\mu &=&{1\over 2}(m_{H_1}^2+m_{H_2}^2+2\mu ^2)\sin 2\beta \;
\end{eqnarray}
For given values of the GUT parameters $m_{1/2}, m_0, A_G$ as well as 
$\tb$, the first minimization equation can be solved for $\mu$ [to within 
a sign]; the second equation can then be solved for $B$. Since $m_{H_1}^2$ 
and $m_{H_2}^2$ are related to $M_A$ through the RGEs, the solution for 
$\mu$ and $B$ can be approximately expressed as a function of $M_A$
and $\tb$.  The power of supergravity models with radiative
electroweak symmetry breaking becomes apparent when one includes the
one-loop contributions to the Higgs potential. It is through these
one--loop terms that most of the supersymmetric particle masses are
determined; the minimization conditions [which are also solved for
$\mu$ to within a sign and $B$] $\it{fix}$ the masses in order that the
electroweak symmetry is broken correctly, {\it i.e.} with the correct
value of $M_Z$. [U(1)$_{\rm EM}$ and SU(3) remain unbroken of course].
The one--loop contributions and the minimization equations are given in
Ref.\cite{R16} to which we refer for details. \s 

A heavy top quark is required to break the electroweak symmetry
radiatively,  since it is the large top Yukawa coupling which will drive
one of the Higgs mass parameters squared to a negative value. As
emphasized before, the additional condition of unification of the $b$--$\tau$
Yukawa couplings gives rise to stringent constraints on $\tb$. The
attractive idea of explaining the large top Yukawa coupling as a result
of a fixed point solution of the RGEs leads to a relationship between
$M_t$ and the angle $\beta$, $M_t \simeq (200~{\rm GeV})\sin \beta$ 
for $\tb \lsim 10$, giving a further constraint on the model. \s 

To limit the parameter space further, one could require that the SUGRA
model is not fine--tuned and the SUSY breaking scale should not be too
high, a constraint which can be particularly restrictive in the small
$\tb$ region. However, the degree of fine--tuning which can be
considered acceptable is largely a matter of taste, so we disregard this
issue in our analysis. \s 

We now detail the calculations of the supersymmetric particle spectrum more
precisely. We incorporate boundary conditions at both electroweak and GUT
scales, adopting the ambidextrous approach of Ref.\cite{R16}. We specify the
values of the gauge and Yukawa couplings at the electroweak scale, in
particular $M_t$, $\tb$ and $\alpha_s$. The gauge and Yukawa couplings are then
evolved to the GUT scale $M_G$ [defined to be the scale $\tilde{\mu}$ for which
$\alpha_1(\tilde{\mu}) = \alpha_2( \tilde{\mu})$] using the two--loop RGEs
\cite{R15}. At $M_G$ we specify the soft supersymmetry breaking parameters
$m_{1/2}$, $m_0$ and $A_G$. We then evolve parameters down to the electroweak
scale where we apply the one--loop minimization conditions derived from the
one--loop effective Higgs potential and solve for $\mu$ to within a sign and
$B$ [we then can integrate the RGEs back to $M_G$ and obtain $\mu_G$ and
$B_G$]. 
By this procedure,  the supersymmetric 
spectrum is
completely specified; that is, we generate a unique spectrum corresponding to
particular values of $m_{1 /2}$, $m_0$, $A_G$, $\tb$ and the sign of $\mu$. It
turns out that the spectrum is nearly independent of $A_G$, for $|A_G| \lsim
500$ GeV. In most of our calculations, we substitute a particular value of
$M_A$ for $m_0$ in order to introduce a mass parameter which can be measured
directly; in this case the value of $m_0$ is chosen [by iteration] so as to produce 
the desired value of $M_A$. \s 

We discuss the SUSY spectrum and its phenomenological
implications for two representative points in the $M_t$--$\tan\beta$
plane\footnote{Our numerical analysis is consistent with the numbers
obtained in Ref.\cite{R35}, once their value of $A_\tau$ in 
Tab.2 is corrected. We thank W. de Boer for a 
discussion on this point.}. We choose $M_t^{\rm pole} = 175$ GeV, 
consistent with the most
recent Tevatron analyses \cite {R12} throughout our calculations, and 
values of $\tb = 1.75$ and 50, which are required (within
uncertainties) by $b$--$\tau$ unification at $M_G$. In particular, we
emphasize the low $\tb$ solutions; they are theoretically favored from
considerations such as $b \rightarrow s\gamma$ \cite{R13} and cosmological
constraints \cite{R17}. The low $\tb$ solutions generate much lighter
SUSY spectra, more likely to be seen at future $e^+ e^-$ colliders. In
both the low and high $\tb$ regions we take\footnote{This 
corresponds to the $\sin^2\theta_W$ value quoted and 
compared with the high--precision
electroweak analyses in the Introduction.} 
$\alpha_s(M_Z^{}) = 0.118$ \cite{R18} and $A_G = 0$, though the
qualitative behavior in each region should not depend greatly on these
parameters. 

\bigskip

\nn (a) \underline{Low $\tan \beta$}

\bigskip

\nn As a typical example of the low $\tan \beta$ region we consider the
point $M^{\rm pole}_t = 175$ GeV and $\tan \beta = 1.75$ for which
$\lambda_t(M_G)$ lies in its ``fixed-point'' region \cite{R11,R19}. If $M_A$
is fixed, the scalar mass parameter $m_0$ can be calculated as a
function of the common gaugino mass parameter $m_{1/2}$ so that
all Higgs and supersymmetric particle masses can in principle be
parameterized by $m_{1/2}$. The correlation between $m_0$ and
$m_{1/2}$ is shown in Fig.2 for three values of $M_A =
300,600$ and 900 GeV in the low $\tb$ region. \s 

Some of the parameter space is already eliminated by experimental bounds
on the light Higgs mass, the chargino/neutralino masses, the light stop
mass, the slepton masses and the squark/gluino masses from LEP1/1.5 and
the Tevatron \cite{R20}. The lower limits are indicated by the 
non--solid lines in Fig.2. Low values of $m_{1/2} \lsim 60$ GeV are excluded
by the lower bound on the gaugino masses. For $\mu>0$, the bound from
the negative search of charginos at LEP1.5 almost rules out completely
the scenario with $M_A \lsim 300$ GeV. If the $h$ boson is not
discovered at LEP2, i.e. if $M_h \gsim 95$ GeV, the whole $\mu<0$
scenario [for $m_{1/2},m_0 < 500$ GeV] can be excluded, 
while for $\mu > 0$ only the $m_{1/2}>200$
GeV range [which implies very large values of $M_A$] would survive. The
requirement that the lightest neutralino is the LSP, and therefore its
mass is larger than the lightest $\tilde{\tau}$ mass, excludes a small
edge of the parameter space [dotted line] at small $m_0$ with $m_{1/
2} > 200$ GeV in the $\mu < 0$ case. \s 

The masses of the Higgs bosons are shown in Fig.3a as a function of
$m_{1/2}$ for $\tb=1.75$, both signs of $\mu$ and for two
representative values of $m_0=100$ and 500 GeV. The lightest Higgs boson
has a rather small tree--level mass and $M_h$ comes mainly from
radiative corrections; the maximal values [for $m_{1/2} \sim 400$ GeV]
are $M_h^{\rm max} \sim 90$
GeV for $\mu<0$ and $\sim 100$ for $\mu>0$. Because the pseudoscalar
mass is approximately given by $M_A^2 \sim B\mu /\sin2\beta \sim
B\mu$ [at the tree--level] and since $B\mu $ turns out to be large in
this scenario, the pseudoscalar $A$ is rather heavy especially for large
values of $m_0$, and thus is almost mass degenerate with the heavy
CP--even and charged Higgs bosons, $M_A \sim M_H \simeq M_{H^\pm}$. Note
that $M_A$ is below the $t\bar{t}$ threshold, $M_A \lsim 350$ GeV, only
if $m_0$ and $m_{1/ 2}$ are both of ${\cal O}(100)$ GeV. \s 

The chargino/neutralino and sfermion masses are shown Fig.3b-d as a
function of $m_{1 /2}$ for the two values $M_A=300$ and 600 GeV and
for both signs of $\mu$. In the case of charginos and neutralinos, the 
masses are related through RGEs by the same ratios that
describe the gauge couplings at the electroweak scale. The LSP is almost
bino--like [with a mass $m_{\chi_1^0} \sim M_1$] while the
next--to--lightest neutralino and the lightest chargino are
wino--like [with masses $m_{\chi_2^0} \sim m_{\chi_1^+} \sim M_2
\sim 2 m_{\chi_1^0}$]. The heavier neutralinos and chargino are
primarily higgsinos with masses $m_{\chi_3^0} \sim m_{\chi_4^0} \sim
m_{\chi_2^+} \sim |\mu|$. The analytical expressions for the chargino 
and neutralino masses [and their limiting values for large $|\mu|$]
are given in Appendix A. Note that the masses approximately scale as 
$M_A$ and that 
the decay of the heavy scalar and pseudoscalar Higgs bosons into 
pairs of the heaviest charginos 
and neutralinos is kinematically not allowed. \s 

The left-- and right--handed charged sleptons and sneutrinos are
almost mass degenerate, the mass differences not exceeding ${\cal
O}(10)$ GeV; the mixing in the $\tau$ sector is rather small 
for small $\tb$, allowing one to
treat all three generations of sleptons on the same footing. In the case of
squarks, only the first two generations are degenerate, with left-- and
right--handed squarks having approximately the same mass. The mixing in
the stop as well as in the sbottom sector leads to a rather substantial
splitting between the two stop or sbottom mass eigenstates. Only for
small values of $M_A$ and for $\mu<0$ is $\tilde{b}_1$ the lightest
squark; otherwise $\tilde{t}_1$ is the lightest squark state. Note that
the squark masses increase with $m_{1 /2}$ and that they scale as $M_A$ 
i.e. as $|\mu|$. The slepton masses decrease with increasing $m_{1/2}$:
this is due to the fact that when $m_{1/2}$ increases and $M_A$
is held constant, $m_0$ decreases 
[see Fig.2], and the dependence of the slepton masses on $m_0$ is 
stronger [for fixed $m_0$, the slepton masses would increase
with increasing $m_{1/2}$]. The analytical 
expressions for the sfermion masses are given in Appendix B.

\bigskip

\nn (b) \underline{High $\tan \beta$}

\bigskip

In this region we take $\tb = 50$ as a representative example, a value
consistent with the unification of the $t$, $b$ and $\tau$ Yukawa 
couplings. The set of possible solutions in the parameter space [$m_{1
/2}$, $m_0$] for $M_A = 300$ and 600 GeV is shown in Fig.4. 
At $\tb = 50$ and $M^{\rm pole}_t = 175$ GeV, we find solutions only for
$\mu <0 $; this is a result of the large one--loop contribution
to $M_A$, the sign of which depends on $\mu$ \cite{R36}. The boundary
contours given in the figure correspond to tachyonic solutions in the
parameter space: $m_{\tilde{\tau_1}}^2 < 0$, $M_A^2 < 0$ or $M^2_h <0$
at the tree--level. The latter constraint is important for algorithmic
reasons: $M^2_h$ at the tree--level enters into the minimization
equations in the form of a logarithm \cite{R16}. Also the
requirement of the lightest neutralino to be the LSP excludes a small
edge of the parameter space at small values of $m_0$; this explains
why the curves for $M_A=300$ and 600 GeV are not extended to low
$m_0$ values. \s

\begin{small}

\vskip 0.3in
{\center \begin{tabular}{|c|c|c|c|c|}
\hline 
\multicolumn{1}{|c|}{Particle}
&\multicolumn{1}{|c|}{Mass (GeV)}
&\multicolumn{1}{|c|}{Mass (GeV)}
&\multicolumn{1}{|c|}{Mass (GeV)}
&\multicolumn{1}{|c|}{Mass (GeV)}
\\[0.2cm] \hline 
\multicolumn{1}{|c|}{$M_A$}
&\multicolumn{1}{|c|}{300}
&\multicolumn{1}{|c|}{300}
&\multicolumn{1}{|c|}{600}
&\multicolumn{1}{|c|}{600}
\\[0.2cm]  \hline 
\multicolumn{1}{|c|}{($m_{1 /2}$, $m_0$)}
&\multicolumn{1}{|c|}{(364,250)}
&\multicolumn{1}{|c|}{(352,800)}
&\multicolumn{1}{|c|}{(603,300)}
&\multicolumn{1}{|c|}{(590,800)}
\\[0.2cm]  \hline \hline 
\multicolumn{1}{|c|}{$\tilde{g}$}
&\multicolumn{1}{|c|}{940}
&\multicolumn{1}{|c|}{910}
&\multicolumn{1}{|c|}{1557}
&\multicolumn{1}{|c|}{1524}
\\[0.2cm]  \hline 
\multicolumn{1}{|c|}{$\tilde{t_1}$,$\tilde{t_2}$}
&\multicolumn{1}{|c|}{662,817}
&\multicolumn{1}{|c|}{753,896}
&\multicolumn{1}{|c|}{1115,1285}
&\multicolumn{1}{|c|}{1156,1325}
\\[0.2cm]  \hline 
\multicolumn{1}{|c|}{$\tilde{b_1}$,$\tilde{b_2}$}
&\multicolumn{1}{|c|}{689,787}
&\multicolumn{1}{|c|}{804,894}
&\multicolumn{1}{|c|}{1159,1260}
&\multicolumn{1}{|c|}{1220,1312}
\\[0.2cm]  \hline 
\multicolumn{1}{|c|}{$\tilde{u_1}$,$\tilde{u_2}$}
&\multicolumn{1}{|c|}{881,909}
&\multicolumn{1}{|c|}{1144,1164}
&\multicolumn{1}{|c|}{1431,1479}
&\multicolumn{1}{|c|}{1586,1628}
\\[0.2cm]  \hline 
\multicolumn{1}{|c|}{$\tilde{d_1}$,$\tilde{d_2}$}
&\multicolumn{1}{|c|}{878,912}
&\multicolumn{1}{|c|}{1142,1167}
&\multicolumn{1}{|c|}{1425,1481}
&\multicolumn{1}{|c|}{1582,1630}
\\[0.2cm]  \hline 
\multicolumn{1}{|c|}{$\tilde{\tau_1}$,
$\tilde{\tau_2}$; $\tilde{\nu_{\tau}}$}
&\multicolumn{1}{|c|}{165,365; 325}
&\multicolumn{1}{|c|}{567,740; 729}
&\multicolumn{1}{|c|}{255,517; 485}
&\multicolumn{1}{|c|}{586,812; 799}
\\[0.2cm]  \hline 
\multicolumn{1}{|c|}{$\tilde{e_1}$, $\tilde{e_2}$; $\tilde{\nu_{e}}$}
&\multicolumn{1}{|c|}{290,360; 351}
&\multicolumn{1}{|c|}{813,838; 835}
&\multicolumn{1}{|c|}{381,519; 513}
&\multicolumn{1}{|c|}{833,901; 898}
\\[0.2cm]  \hline 
\multicolumn{1}{|c|}{$\chi^{\pm}_i$}
&\multicolumn{1}{|c|}{268,498}
&\multicolumn{1}{|c|}{261,536}
&\multicolumn{1}{|c|}{452,764}
&\multicolumn{1}{|c|}{443,779}
\\[0.2cm]  \hline 
\multicolumn{1}{|c|}{$\chi^0_i$}
&\multicolumn{1}{|c|}{144,268,485,496}
&\multicolumn{1}{|c|}{139,261,526,534}
&\multicolumn{1}{|c|}{239,452,754,763}
&\multicolumn{1}{|c|}{234,443,771,778}
\\[0.2cm]  \hline 
\multicolumn{1}{|c|}{$M_A$,$M_{H^{\pm}}$,$M_H$,$M_h$}
&\multicolumn{1}{|c|}{300,315,300,124}
&\multicolumn{1}{|c|}{300,315,300,124}
&\multicolumn{1}{|c|}{600,608,600,131}
&\multicolumn{1}{|c|}{600,608,600,131}
\\[0.2cm]  \hline 
\end{tabular}
\vskip .1in }
\begin{center}
\nn {\bf Tab.1:}  Particle spectra for $M_t^{\rm pole} =$ 175
GeV, $\tan\beta=50$ for selected $M_A, m_{1/2}$ and $m_0$ values.   \\
\end{center}
\vskip .2in

\end{small}

\smallskip

The sparticle spectra for $M_A =$ 300 and 600 GeV and two sets of $m_{1
/2}$ and (extreme) $m_0$ values are shown in Table 1. In all these cases, the
particle spectrum is very heavy; hence most of the SUSY decay channels
of the Higgs particles are shut for large $\tb$. The only allowed decay
channels are $H,A \rightarrow \tilde{\tau_1} \tilde{\tau_1}, \chi^0_1
\chi^0_1$ and $H^\pm \rightarrow \tilde{\tau_1} \tilde{\nu}$ [for large
$M_A$ values]. 
However, the branching ratios of these decay channels are suppressed by
large $b \bar{b}$ and $t \bar{b}$ widths of the Higgs particles for
large $\tb$: while the supersymmetric decay widths are of the order 
${\cal O}(0.1$ GeV), the decays involving $b$ quarks have widths 
${\cal O}(10$ GeV) and dominate by 2 orders of magnitude. 

\bigskip 

\nn (c) \underline{Additional Constraints}

\bigskip

\nn There are additional experimental constraints on the 
parameter space for both high and low $\tb$; the most important
of these are the $b \rightarrow s \gamma$, $Z \rightarrow 
b \overline{b}$, and dark matter [relic LSP abundance] constraints.
These constraints are much more restrictive in the high $\tb$ case.
\s

Recent studies \cite{R13} have indicated that the combination of $b
\rightarrow s \gamma$, dark matter and $m_b$ constraints 
disfavor the high $\tb$ solution for which the $t$, $b$ and $\tau$
Yukawa couplings are equal, in particular the minimal SUSY--SO(10) model
with universal soft-supersymmetry breaking terms at $M_G$. This model
can, however, be saved if the soft terms are not universal [implying a
higgsino--like lightest neutralino], and there exist theoretical motivations
for non--universal soft terms at $M_G$ \cite{R22}. The presently favored
$Z \ra b\bar{b}$ decay width would favor a very low $A$ mass for large 
$\tb$. \s

For low $\tb$, these additional constraints do not endanger the model,
yet they can significantly reduce the available parameter space. In
particular the $Z \rightarrow b \overline{b}$ constraint
favors a light chargino and light stop for small to moderate
values of $\tb$ \cite{R23,R24} so that they could be detected
at LEP2 \cite{R24}. The dark matter constraint essentially places an
upper limit on $m_0$ and $m_{1/2}$ \cite{R26}.
The $b \rightarrow s \gamma$ constraint \cite{R32}, on the other 
hand, is
plagued with large theoretical uncertainties mainly stemming from the
unknown next-to-leading QCD corrections and uncertainties
in the measurement of $\alpha_s(M_Z)$. However, it is consistent with
the low $\tb$ solution and may in the future be useful in determining
the sign of $\mu$ \cite{R33}. 


\setcounter{equation}{0}
\renewcommand{\theequation}{3.\arabic{equation}}

\subsection*{3. Production Mechanisms}

The main production mechanisms of neutral Higgs bosons at $\ee$ 
colliders are the Higgs--strahlung process and pair production,
\begin{eqnarray}
(a) \ \ {\rm Higgs\mbox{-}strahlung} \hspace{1cm} \ee & 
           \ra &  (Z) \ra Z+h/H 
\hspace{5cm} \non \\
(b) \ \ {\rm pair \ production} \hspace{1cm} \ee & \ra & (Z) \ra A+h/H  
\non
\end{eqnarray}
as well as the $WW$ and $ZZ$ fusion processes, 
\begin{eqnarray}
(c) \ \ {\rm fusion \ processes} \hspace{0.8cm} \ \ee & \ra &  \bar{\nu} 
\nu \ (WW) \ra \bar{\nu} \nu \ + h/H \hspace{3.5cm} \non \\
\ee & \ra &  \ee (ZZ) \ra \ee + h/H  \non
\end{eqnarray}
[The ${\cal CP}$--odd Higgs boson $A$ cannot be produced in the
Higgs--strahlung and fusion processes to leading order since it does not
couple to $VV$ pairs.] 
The charged Higgs particle can be pair produced through virtual photon and
$Z$ boson exchange, 
\begin{eqnarray}
(d) \ \ {\rm charged \ Higgs } \hspace{0.8cm} \ \ee & \ra &  \ (\gamma , 
Z^* ) \ \ra \ H^+ H^- \hspace*{3.93cm} \nonumber 
\end{eqnarray}
[For masses smaller than $\sim 170$ GeV, the charged Higgs boson is 
also accessible in top decays, $\ee \ra t\bar{t}$ with $t \ra H^+b$.] \s

The production cross sections\footnote{The complete analytical expressions 
of the cross sections can be found, e.g., in Ref.\cite{R25}. Note that 
in this 
paper there are a few typos that we correct here: in eq.(20), the factor 
92 should replaced by 96; in the argument of the $\lambda$ function of the 
denominator in eq.(21), the parameter $M_A^2$ should be replaced by 
$M_Z^2$; finally, the minus
sign in the interference term in eq.(25) should be replaced by a plus sign.} 
for the neutral 
Higgs bosons are suppressed by mixing angle factors compared to the SM 
Higgs production,
\begin{eqnarray}
\sigma(\ee \ra Zh) \ , \ \sigma(VV \ra h) \ , \ \sigma(\ee \ra AH) \ \ 
\sim \ \sin^2(\beta-\alpha) \\
\sigma(\ee \ra ZH) \ , \ \sigma(VV \ra H) \ , \ \sigma(\ee \ra Ah) \ \ 
\sim \ \cos^2(\beta-\alpha) 
\end{eqnarray}
while the cross section for the charged Higgs particle does not depend 
on any parameter other than $M_{H^\pm}$. \s

In the decoupling limit, $M_A \gg M_Z$, the $HVV$ coupling vanishes, 
while the $hVV$ coupling approaches the SM Higgs value 
\beq
g_{HVV} & = & \cos(\beta-\alpha) \ra  \sin4\beta M_Z^2/2 M_A^2  \ra 0 \\
g_{hVV} & = & \sin(\beta-\alpha)  \ra 1- {\cal O}(M_Z^4/M_A^4) \ \ \ra 1 
\eeq
Hence, the only relevant mechanisms for the production of the heavy Higgs 
bosons in this limit will be the associated pair production (b) and the pair 
production of the charged Higgs particles (d). The cross sections, in the 
decoupling limit and for $\sqrt{s} \gg M_Z$, are given by [we use $M_H \sim 
M_A$]
\beq
\sigma (\ee \ra AH) &=& \frac{G_F^2 M_Z^4}{96 \pi s} (v_e^2+a_e^2)
\beta_A^3 \\
\sigma (\ee \ra H^+H^-) &=& \frac{2G_F^2 M_W^4 s_W^4 }{3 \pi s } \left[
1+ \frac{v_e v_H}{8 s_W^2 c_W^2} + \frac{(a_e^2+ v_e^2)v_H^2}{
256 c_W^4 s_W^4} \right] \beta_{H^\pm}^3
\eeq
where  $\beta_j=(1-4M_j^2/s)^{1/2}$ is the velocity of Higgs bosons, 
 the $Z$ couplings to electrons are given by $a_e=-1, v_e=-1+4\sin^2
\theta_W$, and to the charged Higgs boson by $v_H=-2+4\sin^2\theta_W$;
$s_W^2 = 1-c_W^2 \equiv \sin^2 \theta_W$. The
cross sections for $hA$ and $HZ$ production vanish in the decoupling limit
since they are proportional to $\cos^2 (\beta- \alpha)$. \s

The cross section for the fusion process, $\ee \ra \bar{\nu}_e \nu_e H$, is
enhanced at high energies since it scales like $M_W^{-2}\log s/M_H^2$. 
This mechanism
provides therefore a useful channel for $H$ production in the mass range of a
few hundred GeV below the decoupling limit and small values of $\tb$, where
$\cos^2(\beta-\alpha)$ is not prohibitively small; the cross section, though,
becomes gradually less important for increasing $M_H$ and vanishes in the
decoupling limit. In the high energy regime, the $WW\ra H$ 
fusion cross section
is well approximated by the expression 
\beq
\sigma( \ee \ra \bar{\nu}_e \nu_e H ) = \frac{G_F^3 M_W^4}{4 \sqrt{2}\pi^3} 
\left[ \left(1+\frac{M_H^2}{s} \right) \log \frac{s}{M_H^2} -2 \left(1-
\frac{M_H^2}{s} \right) \right] \cos^2(\beta-\alpha) 
\eeq
obtained in the effective longitudinal $W$ approximation. Since the NC
couplings are small compared to the CC couplings, the cross section for the
$ZZ$ fusion process is $\sim 16\cos^4 \theta_W$, {\it i.e.} one order of
magnitude smaller than for $WW$ fusion. \s 

Numerical results for the cross sections are shown in Fig.5 at high--energy 
$\ee$ colliders as a function of $\sqrt{s}$ TeV for the two values
$\tb=1.75$ and 50, and for pseudoscalar masses $M_A=300$, 600 and 900 
GeV [note that $M_H \simeq M_{H^\pm} \simeq M_A$ as evident from Figs. 1 
and 3a]. For a luminosity of
$\int {\cal L}=200$ fb$^{-1}$, typically a sample of about $1000$ $HA$
and $H^+H^-$ pairs are predicted for heavy Higgs masses of $\sim 500$
GeV at $\sqrt{s}=1.5$ TeV. For small $\tb$ values, $\tb \lsim 2$, a few 
hundred events are predicted in the $WW \ra H$ fusion process for $H$ 
masses $\sim 300$ GeV. The cross sections for the $hA$ and $HZ$ processes 
are too low, less than $\sim 0.1$ fb, to be useful for $M_H \gsim 300$ GeV; 
Fig.5b. \s 

Note that the cross sections for the production of the lightest Higgs
boson $h$ in the decoupling limit and for $\sqrt{s} \gg M_Z, M_h$ are 
simply given by
\beq
\sigma (\ee \ra Zh) &\simeq& \frac{G_F^2 M_Z^4}{96 \pi s} (v_e^2+a_e^2) \\
\sigma( \ee \ra \bar{\nu}_e \nu_e h ) &\simeq& \frac{G_F^3 M_W^4}{4 \sqrt{2}
\pi^3} \log \frac{s}{M_h^2} 
\eeq
The cross sections are the same as for the SM Higgs particle and 
are very large $\sim 100$ fb, 
especially for the $WW$ fusion mechanism.

\setcounter{equation}{0}
\renewcommand{\theequation}{4.\arabic{equation}}

\subsection*{4. Decay Modes}

\noindent {\bf 4.1 Decays to standard  particles }

\bigskip

For large $\tb$ the Higgs couplings to down--type fermions dominate over
all other couplings. As a result, the decay pattern is in general very
simple. The neutral Higgs bosons will decay into $b\bar{b}$ and
$\tau^+\tau^-$ pairs for which the branching ratios are close to $\sim
90$~\% and $\sim 10$~\%, respectively. The charged Higgs particles decay
into $\tau\nu_{\tau}$ pairs below and into $tb$ pairs above the
top--bottom threshold. \s 

The partial decay widths of the neutral Higgs bosons\footnote{We refrain
from a discussion of the $h$ decays which become SM--like in the
decoupling limit. In addition, we discuss only the dominant two--body
decay modes of the heavy Higgs bosons; for an updated and more detailed
discussion, including also three--body decays, see Ref.\cite{R28}.}, $\Phi=H$
and $A$, to fermions are given by \cite{R4} 
\beq
\Gamma(\Phi \ra \bar{f}f) = N_c \frac{G_F M_\Phi }{4 \sqrt{2} \pi} 
g^2_{\Phi ff}m_f^2 \beta_f^p
\eeq
with $p=3(1)$ for the CP--even (odd) Higgs bosons;  $\beta_f=(1-4m_f^2/
M_\Phi^2) ^{1/2}$ is the velocity of the fermions in the final state,
$N_c$ the color factor. For the decay widths to quark  pairs, the QCD 
radiative corrections are large and must be included; for a recent 
update and a more detailed discussion, see Ref.\cite{R27}. \s

The couplings of the MSSM neutral Higgs bosons [normalized to the SM Higgs 
coupling $g_{H_{\rm SM}ff} = \left[ \sqrt{2} G_F \right]^{1/2} m_f$
and $g_{H_{\rm SM}VV} = 2 \left[ \sqrt{2} G_F \right]^{1/2} M_V^2$] are given 
in Table 2. 
\vspace*{3mm}

\begin{center}
\begin{tabular}{|c||c|c||c|} \hline
& & & \s
$\hspace{1cm} \Phi \hspace{1cm} $ &$ g_{ \Phi \bar{u} u} $ & $
g_{\Phi \bar{d} d} $ & $ g_{\Phi VV} $ \\
& & & \\ \hline \hline
& & & \\ 
$h$  & \ $\; \cos\alpha/\sin\beta       \; $ \ & \ $ \; -\sin\alpha/
\cos\beta \; $ & $ \sin(\beta-\alpha) $ \\
$H$  & \       $\; \sin\alpha/\sin\beta \; $ \ & \ $ \; \cos\alpha/
\cos\beta \; $ & $ \cos(\beta-\alpha) $ \\
$A$  & \ $\; 1/ \tb \; $        \ & \ $ \; \tb \; $ & $0$ \\[0.3cm] \hline
\end{tabular}
\end{center}

\vspace*{3mm}

\nn {\small {\bf Tab.~2}: Higgs boson couplings in the MSSM to fermions 
and gauge bosons relative to the SM Higgs  couplings.}

\vspace*{4mm}

In the decoupling limit, $M_A \gg M_Z$, we have [at tree--level] 
\beq
\cos\alpha & \sim &
\sin\beta -\cos\beta \sin4\beta \frac{M_Z^2}{2M_A^2} 
\ra \sin\beta \\
\sin\alpha& \sim&
{} -\cos\beta+\sin\beta \sin4\beta \frac{M_Z^2}{2M_A^2}
\ra {}-\cos\beta
\eeq
Therefore the $hff$ couplings reduce to the SM 
Higgs couplings, while the $Hff$ couplings become equal
to those of the pseudoscalar boson $A$,
\beq
\cos\alpha/ \sin\beta & \ra &  \ 1 \non \\
-\sin\alpha/ \cos\beta & \ra & \ 1 \non \\
-\sin\alpha/\sin\beta & \ra & \ 1/\tb \non \\
\cos\alpha/\cos\beta &\ra &  \ \tb 
\eeq

The partial width of the decay mode $H^+\ra u \bar{d}$ is given by
\beq
\Gamma(H^+\ra u\bar{d}) &=& \frac{N_c G_F}{4\sqrt{2} \pi} \frac{\lambda^{1/2}
_{ud,H^{\pm} } } {M_{H^\pm}} \, |V_{ud}|^2 \times \non \\
& & \left[ (M_{H^\pm}^{2} -m_{u}^{2}-m_{d}^{2}) \left( m_{d}^{2} {\tg}^2
\beta + m_u^2{\rm ctg}^2 \beta \right) -4m_u^2m_{d}^2 \right]
\end{eqnarray}
with  $V_{ud}$ the CKM--type matrix element for quarks and
$\lambda$ is the two--body phase space function defined by
\beq
\lambda_{ij,k} = (1-M_i^2/M_k^2- M_j^2/M_k^2)^2-4M_i^2 M_j^2/M_k^4 
\eeq
For decays into quark pairs, the QCD corrections must be also included. \s

Below the $\bar{t}t$ threshold, a variety of channels is open
for the decays of the heavy CP--even Higgs bosons, the most important
being the cascade decays $H \ra \Phi \Phi$ with $\Phi=h$ or $A$, with
a partial width [for real light Higgs bosons]
\begin{eqnarray}
\Gamma(H \ra \Phi \Phi) = \frac{G_F}{16\sqrt{2} \pi} \frac{M_Z^4}{M_H}
 g^2_{H\Phi \Phi} \beta_{\Phi}
\label{HAA}
\end{eqnarray}
where $\beta_{\Phi}=(1-4 M_{\Phi}^2/M_H^2)^{1/2}$ and 
the radiatively corrected three--boson self--couplings [to leading 
order], in units of $g_Z'=(\sqrt{2}G_F)^{1/2}M_Z^2$, are given by 
\begin{eqnarray}
g_{Hhh} &=& 2 \sin 2\alpha \sin (\beta+\alpha) -\cos 2\alpha \cos(\beta
+ \alpha) + 3 \frac{\epsilon}{M_Z^2} \frac{\sin \alpha \cos^2\alpha }{\sin
\beta}
\label{3h2}\\
g_{HAA} &=& - \cos 2\beta \cos(\beta+ \alpha)+ \frac{\epsilon}{M_Z^2}
\frac{\sin \alpha \cos^2\beta }{\sin \beta} \label{3h3} \non 
\end{eqnarray}
In contrast to the previous couplings, the leading $m_t^4$ radiative
corrections cannot be absorbed entirely in the redefinition of the
mixing angle $\alpha$, but they are renormalized by an explicit term
depending on the parameter $\epsilon$ given by [$M_S$ is the common
squark mass at the electroweak scale]
\begin{eqnarray}
\epsilon = \frac{3 G_F}{\sqrt{2} \pi^2} \frac{m_t^4}{ \sin^2\beta} \, \log 
\left( 1+ \frac{M_S^2}{m_t^2} \right)
\end{eqnarray}
In the decoupling limit, these couplings approach the values
\beq
g_{Hhh} & \ra & \frac{3}{2}\sin4\beta \non \\
g_{HAA} & \ra & -\frac{1}{2}\sin4\beta
\eeq

In the mass range above the $WW$ and $ZZ$ thresholds, where the $HVV$ 
couplings
are not strongly suppressed for small values of $\tb$, the partial widths of
the $H$ particle into massive gauge bosons can also be substantial; they are
given by 
\begin{eqnarray}
\Gamma (H \ra VV) =  \frac{ \sqrt{2} G_F \cos^2 (\alpha-\beta)} {32 \pi}
M_H^3 (1-4\kappa_V+12\kappa_V^2) (1-4\kappa_V)^{1/2} \ \delta_V'
\end{eqnarray}
with $\kappa_V=M_V^2/M_H^2$ and $\delta_V'=2(1)$ for  $V=W(Z)$. \s

For small values of $\tb$ and below the $\bar{t}t$ and the $t\bar{b}$ 
thresholds, the pseudoscalar and charged Higgs bosons can decay
into the lightest Higgs boson $h$ and a gauge boson; however these
decays are suppressed by $\cos^2(\beta-\alpha)$ and therefore are 
very rare for large $A$ masses. The partial decay widths are given by
\begin{eqnarray} 
\Gamma(A \ra hZ) &=& \frac{G_F \cos^2 (\beta-\alpha) }{8\sqrt{2} \pi} 
\ \frac{M_Z^4}{M_A} \lambda^{1/2}_{Zh,A} \lambda_{Ah,Z} \non \\
\Gamma(H^{+} \rightarrow hW^+) &=& 
\frac{G_F \cos^2 (\beta-\alpha) }{8\sqrt{2}\pi} \frac{M_W^4}{M_{H^\pm}} 
\lambda^{\frac{1}{2}}_{Wh,H^{\pm}} \lambda_{H^{\pm}h,W} 
\hspace*{0.2cm}
\end{eqnarray}

In the decoupling limit, the partial widths of all decays of the
heavy CP--even, CP--odd and charged Higgs bosons involving gauge bosons
vanish since $\cos^2(\beta-\alpha)\ra0$. In addition, the $H \ra hh$  
decay width is very small
since it is inversely proportional to $M_H$, and $H\ra AA$ is not 
allowed kinematically. Therefore, the only
relevant channels are the decays into $\bar{b}b/\bar{t}t$ for the
neutral and $t\bar{b}$ for the charged Higgs bosons. The total decay widths of
the three bosons $H,A$ and $H^\pm$, into standard particles can be 
approximated in this limit by 
\beq
\Gamma(H_k \ra {\rm all}) = \frac{3 G_F}{4\sqrt{2} \pi} M_{H_k} 
\left[ m_{b}^{2} {\tg}^2\beta + m_t^2{\rm ctg}^2 \beta \right]
\eeq 
[We have neglected the small contribution of the decays into $\tau$ leptons 
for large $\tb$.] 

\vspace{0.3in}

\noindent {\bf 4.2 Decays to charginos and neutralinos}

\vspace{0.3in}

The decay widths of the Higgs bosons $H_k$  [$k=(1,2,3,4)$ correspond
to $ (H,h,A,H^\pm)$] into neutralino and chargino pairs are given by 
\cite{R29}
\begin{eqnarray}
\Gamma (H_k \ra \chi_i \chi_j) = \frac{G_F M_W^2}{2 \sqrt{2} \pi}
\frac{ M_{H_k} \lambda_{ij,k}^{1/2} }{1+\delta_{ij}} \hspace*{-3mm}
&& \left[ (F_{ijk}^2 + F_{jik}^2) \left(1- \frac{ m_{\chi_i}^2}{M_{H_k}^2}
        - \frac{ m_{\chi_j}^2}{M_{H_k}^2} \right) \right. \non \\
&& \left. -4\eta_k \epsilon_i \epsilon_j F_{ijk} F_{jik} \frac{ m_{\chi_i} 
m_{\chi_j}} {M_{H_k}^2} \right]
\end{eqnarray}
where $\eta_{1,2,4}=+1$, $\eta_3=-1$ and $\delta_{ij}=0$ unless the
final state consists of two identical (Majorana) neutralinos in which
case $\delta_{ii}=1$; $\epsilon_i =\pm 1$ stands for the sign of the 
$i$'th eigenvalue of the neutralino mass matrix [the matrix $Z$ is defined 
in the convention of Ref.\cite{R2a}, and the eigenvalues of the mass matrix 
can be either positive or negative] while  $\epsilon_i=1$ for charginos;
$\lambda_{ij,k}$ is the usual two--body phase 
space function given in eq.(4.6).\s 

In the case of neutral Higgs boson decays, the coefficients $F_{ijk}$ are
related to the elements of the matrices $U,V$ for charginos and
$Z$ for neutralinos,
\begin{eqnarray}
H_k \ra \chi_i^+ \chi_j^- \ &:& F_{ijk}= \frac{1}{\sqrt{2}} \left[
e_k V_{i1}U_{j2} -d_k V_{i2}U_{j1} \right] \\
H_k \ra \chi_i^0 \chi_j^0 \ \ &:& F_{ijk}= \frac{1}{2}
\left( Z_{j2}- \tan\theta_W
Z_{j1} \right) \left(e_k Z_{i3} + d_kZ_{i4} \right) \ + \ i \leftrightarrow j
\end{eqnarray}
with the coefficients $e_k$ and $d_k$ given by
\begin{eqnarray}
e_1/d_1=\cos\alpha/-\sin \alpha \ , \
e_2/d_2=\sin\alpha/\cos \alpha \ ,  \
e_3/d_3=-\sin\beta/\cos \beta
\end{eqnarray}
For the charged Higgs boson, the coupling to neutralino/chargino
pairs are  given by  
\begin{eqnarray}
F_{ij4} &  = & \cos\beta \left[ Z_{j4} V_{i1} + \frac{1}{\sqrt{2}}
\left( Z_{j2} + \tan \theta_W Z_{j1} \right) V_{i2} \right] \non \\
F_{ji4} & = & \sin \beta \left[ Z_{j3} U_{i1} - \frac{1}{\sqrt{2}}
\left( Z_{j2} + \tan \theta_W Z_{j1} \right) U_{i2} \right]
\end{eqnarray}
The matrices $U,V$ for charginos and $Z$ for neutralinos
are given in Appendix A. \s

Since in most of the parameter space discussed in Section 2, the
Higgs--higgsino mass parameter $|\mu|$ turned out to be very large, $|\mu| \gg
M_1 , M_2 , M_Z$, it is worth discussing the Higgs decay widths into charginos
and neutralinos in this limit. First, the decays of the neutral Higgs bosons
into pairs of [identical] neutralinos and charginos $H_k \ra \chi_i \chi_i$
will be suppressed by powers of $M_Z^2/\mu^2$. This is due to the fact that 
neutral Higgs bosons mainly couple to {\it mixtures} of higgsino and gaugino 
components, and in the large $\mu$ limit, neutralinos and charginos are either
pure higgsino-- or pure gaugino--like. For the same reason, decays $H^+ \ra
\chi_{1,2}^0 \chi_1^+$ and $\chi_{3,4}^0 \chi_2^+ $ of the charged Higgs 
bosons
will be suppressed. Furthermore, since in this case 
$M_A$ is of the same order as $|\mu|$, decays into pairs of heavy charginos 
and neutralinos will be kinematically forbidden. Therefore, the channels 
\beq
H,A & \ra & \chi_1^0 \, \chi_{3,4}^0  \ , \ \chi_2^0 \, \chi_{3,4}^0  \ \ 
{\rm and} \ \chi_1^\pm \, \chi_2^\mp \non \\
H^+ & \ra & \chi_1^+ \, \chi_{3,4}^0  \ \ {\rm and} 
\ \chi_2^+ \chi_{1,2}^0 
\eeq
will be the dominant decay channels of the heavy Higgs particles. Up to 
the phase space suppression [i.e. for $M_{A}$ sufficiently larger 
than $|\mu|$], the partial widths of these decay channels, in units of 
$G_F M_W^2 M_{H_k}/( 4 \sqrt{2} \pi)$, are given by \cite{R29}
\beq
\Gamma( H \ra \chi_1^0 \chi_{3,4}^0)& =& 
\frac{1}{2} {\rm tan}^2 \theta_W ( 1 \pm \sin 2\beta) \non \\
\Gamma( H \ra \chi_2^0 \chi_{3,4}^0)&=& 
 \frac{1}{2} (1\pm \sin 2\beta) \non \\ 
\Gamma( H \ra \chi_1^\pm \, \chi_{2}^\mp) & =& 1 \\
\Gamma( A \ra \chi_1^0 \chi_{4,3}^0)  &=&
\frac{1}{2} {\rm tan}^2 \theta_W ( 1 \pm \sin 2\beta) \non \\
\Gamma( A \ra \chi_2^0 \chi_{4,3}^0)  &=& 
\frac{1}{2} (1\pm \sin 2\beta) \non \\
\Gamma( A \ra \chi_1^\pm \, \chi_{2}^\mp)  &=& 1 \\
\Gamma( H^+ \ra \chi_1^+ \, \chi_{3,4}^0) & =& 1 \non \\
\Gamma( H^+ \ra \chi_2^+ \, \chi_{1}^0) &=&  {\rm tan}^2 \theta_W  \non \\
\Gamma( H^+ \ra \chi_2^+ \, \chi_{2}^0)  &=& 1 
\eeq
[We have used the fact that in the decoupling limit $\sin 2\alpha=
-\sin 2\beta$.] If all these channels are kinematically allowed, the total
decay widths of the heavy Higgs bosons to chargino and neutralino
pairs will be given by the expression
\begin{eqnarray}
\Gamma (H_k \ra \sum \chi_i \chi_j) = \frac{3 G_F M_W^2}{4 \sqrt{2} \pi}
M_{H_k} \left( 1+\frac{1}{3} \tan^2 \theta_W \right)
\end{eqnarray}
which holds universally for all the three Higgs bosons $H,A,H^\pm$. 

\bigskip 

\noindent {\bf 4.3 Decays to squarks and sleptons}

\bigskip

\nn Decays of the neutral and charged Higgs bosons, $H_k=h,H,A,H^\pm$,
to sfermion pairs can be written as 
\begin{eqnarray}
\Gamma (H_k \ra \tilde{f}_i \tilde{f}_j ) = \frac{N_C G_F }{2 \sqrt{2} 
\pi M_{H_k} } \, \lambda^{1/2}_{ \tilde{f}_i \tilde{f}_j, H_k} \, 
g_{H_k \tilde{f}_i \tilde{f}_j}^2
\end{eqnarray}
$\tilde{f}_{i}$ with $i=1,2$ are the sfermion mass eigenstates
which are related to the current eigenstates $\tilde{f}_{L}, \tilde{f}_{R}$ 
by
\beq
\tilde{f}_1 &=& \ \tilde{f}_L \cos\theta_f + \tilde{f}_R \sin \theta_f 
\non \\
\tilde{f}_2 &=& -\tilde{f}_L \sin\theta_f + \tilde{f}_R \cos \theta_f 
\eeq
The mixing angles $\theta_f$ are proportional to the masses of the partner 
fermions and they are important only in the case of third generation 
sfermions. The couplings $g_{H_k \tilde{f}_i \tilde{f}'_j}$ of the neutral 
and charged Higgs bosons $H_k$ to sfermion mass eigenstates are 
superpositions of the couplings of the current eigenstates, 
\beq
g_{H_k \tilde{f}_i \tilde{f}'_j} = \sum_{\alpha, \beta=L,R} T_{ij \alpha 
\beta} \ g_{\Phi \tilde{f}_\alpha \tilde{f}'_\beta} 
\eeq
The elements of the $4 \times 4$ matrix $T$ are given in Tab.3a. 
The couplings $g_{H_k \tilde{f}_\alpha \tilde{f}'_\beta}$, in the 
current eigenstate basis $\tilde{f}_{\alpha,\beta}=
\tilde{f}_{L,R}$ [normalized to $2 (\sqrt{2} G_F)^{1/2}$] 
may be written as \cite{R4,R29}
\beq
g_{H_k \tilde{f}_L \tilde{f}_L} &=& m_f^2 g^{\Phi }_1 + M_Z^2 (T_3^f 
-e_f s_W^2) g_2^{\Phi } \non \\
g_{H_k \tilde{f}_R \tilde{f}_R} &=& m_f^2 g^{\Phi 
}_1 + M_Z^2 e_f s_W^2 g_2^{\Phi } \non \\
g_{H_k \tilde{f}_L \tilde{f}_R} &=& - \frac{1}{2} m_f 
\left[ \mu g^{\Phi }_3 - A_f g_4^{\Phi } \right] 
\eeq
for the neutral Higgs bosons, $H_k=h,H,A$. $T_3=\pm 1/2$ is the isospin 
of the [left--handed] sfermion and $e_f$ its 
electric charge. The coefficients 
$g_i^{\Phi}$ are given in Tab.3b; in the decoupling limit, the coefficients
$g_2^{\Phi}$ reduce to 
\beq
\cos(\beta+\alpha) & \ra & \sin2\beta \non \\
\sin(\beta+\alpha) & \ra & -\cos2\beta 
\eeq
[for the other coefficients, see eqs.(4.2)].
For the charged Higgs bosons, the couplings [also normalized to $2 (\sqrt{2}
G_F )^{1/2}$] are 
\beq
g_{H^+ \tilde{u}_\alpha \tilde{d}_\beta} &=& - \frac{1}{\sqrt{2}}
\left[  g_1^{\alpha \beta} + M_W^2 g_2^{\alpha \beta} \right]
\eeq
with the coefficients $g_{1/2}^{\alpha \beta}$ with $\alpha, \beta=L,R$ 
listed in Table 3c. \s

\bigskip

\begin{center}
\begin{tabular}{|c|cccc|} \hline
&&&& \\
$i,j \ /  \ \alpha, \beta$ & $ \mbox{\small LL} $ & $
\mbox{\small RR}$ &  $\mbox{\small LR} $ & $\mbox{\small RL}$ \\ 
&&&& \\ \hline 
&&&&  \\
11 & $ \cos \theta_f \cos \theta_{f'} $
   & $ \sin \theta_f \sin \theta_{f'} $
   & $ \cos \theta_f \sin \theta_{f'} $
   & $ \sin \theta_f \cos \theta_{f'} $ \\ 
12 & $ -\cos \theta_f \sin \theta_{f'}$
   & $\sin \theta_f \cos \theta_{f'}$
   & $\cos \theta_f \cos \theta_{f'}$
   & $-\sin \theta_f \sin \theta_{f'}$ \\ 
21 & $-\sin \theta_f \cos \theta_{f'}$
   & $\cos \theta_f \sin \theta_{f'}$
   & $-\sin \theta_f \sin \theta_{f'}$
   & $\cos \theta_f \cos \theta_{f'} $ \\ 
22 & $\sin \theta_f \sin \theta_{f'}$
   & $\cos \theta_f \cos \theta_{f'}$
   & $-\sin \theta_f \cos \theta_{f'} $
   & $-\cos \theta_f \sin \theta_{f'} $\\ 
&&&& \\ \hline
\end{tabular}
\end{center}

\nn {\small {\bf Tab.~3a}: Transformation matrix for the Higgs couplings to 
sfermions in the presence of mixing.}

\bigskip

\begin{center}
\begin{tabular}{|c|c|c|c|c|c|} \hline
& & & & & \\
$ \ \ \tilde{f} \ \ $ & $\ \ \Phi \ \ $ & $\ \ g_1^{\Phi } \ \ $ 
& $g_2^{\Phi }$ & $g_3^{\Phi }$ & $g_4^{\Phi } $ \\ 
&&&&& \\ \hline  &&&&& \\
&$h$ & $\cos\alpha/ \sin \beta $ & $-\sin(\alpha+\beta)$ 
               & $-\sin\alpha/\sin\beta$ & $\cos \alpha/ \sin\beta$ \\
$ \tilde{u}$& $H$ & $\sin\alpha/\sin\beta$ & $\cos(\alpha+\beta)$ & 
    $\cos \alpha /\sin \beta$ & $\sin \alpha/ \sin\beta$ \\ 
& $A$ & $0$ & $0$ & $1$ & $-1/\tb$ \\ 
&&&&& \\
& $h$ & $-\sin\alpha/ \cos \beta $ & $-\sin(\alpha+\beta)$ 
               & $\cos\alpha/\cos\beta$ & $-\sin \alpha/ \cos\beta$ \\
$\tilde{d}$ & $H$ & $\cos\alpha/\cos\beta$ & $\cos(\alpha+\beta)$ & 
    $\sin \alpha /\cos \beta$ & $\cos \alpha/ \cos\beta$ \\ 
&    $A$ & $0$ & $0$ & $1$ & $-\tb$ \\ 
&&&&& \\ \hline
\end{tabular}
\end{center}

\nn {\small {\bf Tab.~3b}: Coefficients in the couplings of 
neutral Higgs bosons to sfermion pairs. } 

\bigskip

\begin{small}
\begin{center}
\begin{tabular}{|c|c|c|c|} \hline
&&& \\
$g^{LL}_{1/2} $ & $g^{RR}_{1/2} $  & $g^{LR}_{1/2} $ & $g^{RL}_{1/2} $ \s
&&& \\ \hline
&&& \\
$ m_u^2/{\rm tan} \beta + m_d^2 {\rm tan} \beta$ &
$ m_u m_d (\tb + 1/ \tb )$ & $m_d (\mu + A_d \tb )$ &
$ m_u( \mu +  A_u / \tb )$ \\
$- \sin 2\beta $ & 0 &0 &0 \\
&&& \\ \hline
\end{tabular}
\end{center}
\end{small}

\nn {\small {\bf Tab.~3c}: Coefficients in the couplings of 
charged Higgs bosons to sfermion pairs. } 

\bigskip

Mixing between sfermions occurs only in the third--generation sector. 
For the first two generations the decay pattern is rather simple. In 
the limit of massless fermions, the pseudoscalar Higgs boson does 
not decay into sfermions since by virtue of CP--invariance it couples 
only to pairs of left-- and right--handed sfermions with the coupling 
proportional to $m_f$. In the asymptotic regime, where the masses 
$M_{H,H^\pm}$ are 
large, the decay widths of the heavy CP--even and charged \cite{Bartl} 
Higgs bosons 
into sfermions are proportional to 
\beq
\Gamma (H,H^+ \ra \tilde{f} \tilde{f} ) \sim \frac{G_F M_W^4}{M_{H}}
\, \sin^2 2 \beta
\eeq
These decay modes can be significant only for low values
$\tb$ [which implies $\sin^2 2 \beta \sim 1$]. However, in this regime 
the decay widths are inversely proportional to $M_H$, and thus cannot 
compete with the decay widths into charginos/neutralinos and ordinary fermions
which increase with increasing Higgs mass. Therefore, the decays
into first and second generations are unlikely to be important. \s

In the case of the third generation squarks, the Higgs decay widths can be
larger by more than an order of magnitude. For instance the decay widths 
of the heavy neutral Higgs boson into top squarks of the same helicity 
is proportional to 
\beq
\Gamma (H \ra \tilde{t} \tilde{t} ) \sim \frac{G_F m_t^4 }{M_{H} {\rm 
tan}^2 \beta } 
\eeq
in the asymptotic region, and it will be enhanced by large coefficients 
[for small $\tb$] compared to first/second generation 
squarks. Conversely, the decay widths into sbottom quarks can be very large 
for large $\tb$. Furthermore, the decays of heavy neutral CP--even 
and CP--odd Higgs bosons to top squarks of different helicities will 
be proportional in the asymptotic region [and for the CP--even, up to 
the suppression by mixing angle] to 
\beq
\Gamma (H,A \ra \tilde{t} \tilde{t} ) \sim \frac{G_F m_t^2 }{M_{H}} \left[ 
\mu + A_t/ {\rm tan} \beta \right]^2 
\eeq
For $\mu$ and $A_t$ values of the order of the Higgs boson masses, these 
decay widths will be competitive with the chargino/neutralino and standard
fermion decays. Therefore, if kinematically allowed, these decay modes
have to be taken into account. 

\bigskip

\noindent {\bf 4.4 Numerical results}

\bigskip

The decay widths of the $H,A$ and $H^\pm$ Higgs bosons into the sum of
charginos and neutralinos, squark or slepton final states, as well as the
standard and the total decay widths are shown in Figs.6a, 7a and 8a 
as a function of
$m_{1/2}$ for two values of the pseudoscalar Higgs boson mass $M_A=300$
and $600$ GeV, and for positive and negative $\mu$ values; $\tb$ is
fixed to $1.75$. \s 

Fig.6a shows the various decay widths for the heavy CP--even Higgs
boson. For $M_A=300$ GeV, the $H\ra t\bar{t}$ channel is still closed
and the decay width into standard particles is rather small, being of
${\cal O}(250)$ MeV. In this case, the decays into the lightest stop
squarks which are kinematically allowed for small values of $m_{1/2}$ 
will be by far the dominant decay channels. This occurs in most of the
$m_{1/2}$ range if $\mu>0$, but if $\mu<0$ only for $m_{1/2} \lsim 50$ GeV 
which is already ruled out by CDF and LEP1.5 data. \s 

The decays into charginos and neutralinos, although one order of magnitude
smaller than stop decays when allowed kinematically, are also very important.
They exceed the standard decays in most of the $m_{1/2}$ range, except for
large values of $m_{1/2}$ and $\mu <0$  where no more phase space 
is available for the Higgs boson to decay into combinations of the heavy 
and light chargino/neutralino states. For small $m_{1/2}$ values, chargino and 
neutralino decays can be larger than the standard decays by up to an 
order of magnitude. \s

As expected, the decay widths into sleptons are rather small and they never
exceed the widths into standard particles, except for large values of 
$m_{1/2}$.
Note that due to the isospin and charge assignments, the coupling of the $H$
boson to sneutrinos is approximately a factor of two 
larger than the coupling to
the charged sleptons. Since all the sleptons of the three generations are
approximately mass degenerate [the mixing in the $\tilde{\tau}$ sector is very
small for low values of $\tb$], the small decay widths into sleptons 
are given by the approximate relation: 
$\Gamma (H \ra \tilde{\nu} \tilde{\nu}) \simeq 4 \Gamma (H \ra \tilde{l}_L
\tilde{l}_L) \simeq 4\Gamma( H \ra \tilde{l}_R \tilde{l}_R)$. \s 

For larger values of $M_H$, $M_H \gsim 350$ GeV, the decay widths into
supersymmetric particles have practically the same size as discussed
previously. However, since the $H\ra t\bar{t}$ channel opens up, the decay
width into standard model particles becomes rather large, ${\cal O}(10$ GeV),
and the supersymmetric decays are no longer dominant. For $M_H\simeq 600$
GeV, Fig.6a, only the $H \ra \tilde{q} \tilde{q}$ decay width can be larger
than the decay width to standard particles; this occurs in the lower range of
the $m_{1/2}$ values. The chargino/neutralino decays have a branching ratio of
$\sim 20\%$, while the branching ratios of the decays into sleptons are below
the $1\%$ level. \s

Fig.6b and 6c show the individual decay widths of the heavy $H$ boson with a
mass $M_H \simeq 600$ GeV into charginos, neutralinos, stop quarks
and sleptons for the set of parameters introduced previously. For decays into
squarks, only the channels $H\ra \tilde{t}_1 \tilde{t}_1, \tilde{t}_1
\tilde{t}_2$, and in a very small range of $m_{1/2}$ values the channel
$H\ra \tilde{b}_1 \tilde{b}_1$, are allowed kinematically [see Fig.3c].
The decay into two different stop states is suppressed by the [small]
mixing angle, and due to the larger phase space the decay
$H \ra \tilde{t}_1 \tilde{t}_1$ is always dominating. \s 

For the decays into chargino and neutralinos, the dominant channels are decays
into mixtures of light and heavy neutralinos and charginos, in particular $H
\ra \chi_1^+ \chi_2^-$ and $H \ra \chi_1^0 \chi_3^0$ or $\chi_2^0 \chi_3^0$.
This can be qualitatively explained, up to phase space suppression factors,  
by recalling the approximate values of the relative branching ratios in 
the large $|\mu|$ limit given in eqs.(4.18--20):
$\Gamma( H \ra \chi_1^\pm \chi_2^\mp) \sim 1$, while $ \Gamma(H \ra \chi_2^0
\chi_3^0) \sim 1$ and $ \Gamma(H \ra \chi_1^0 \chi_3^0) \sim \tan^2 \theta_W$
because $\sin 2\beta$ is close to one.  The mixed decays involving $\chi_4^0$
are suppressed since they are proportional to $(1-\sin 2\beta)$, and all 
other decay channels are suppressed by powers of $M_Z^2/\mu^2$ for large
$|\mu|$ values. \s

The decay widths for the pseudoscalar Higgs boson are shown in Fig.7a. There
are no decays into sleptons, since the only decay allowed by CP--invariance, $A
\ra \tilde{\tau}_1 \tilde{\tau}_2$, is strongly suppressed by the very small
$\tilde{\tau}$ mixing angle. For $M_A=300$ GeV, the decay into the two stop
squark eigenstates, $A\ra \tilde{t}_1 \bar{\tilde{t}}_2$, is not allowed
kinematically and the only possible supersymmetric decays are the decays into
charginos and neutralinos. The sum of the decay widths into these states can be
two orders of magnitude larger than the decay width into standard particles. \s

For values of $M_A$ above the $t\bar{t}$ threshold, the decay width into
charginos and neutralinos is still of the same order as for low $M_A$, but
because of the opening of the $A \ra t\bar{t}$ mode, the total decay width
increases dramatically and the chargino/neutralino decay branching ratio drops
to the level of $20\%$. As in the case of the heavy CP--even Higgs boson $H$,
the relative decay widths of the pseudoscalar boson into charginos and
neutralinos, Fig.7b, are larger for the channels involving mixtures of light
and heavy neutralinos or charginos; the dominant decay modes are, roughly,
$A\ra \chi_1^+ \chi_2^-$ and $A \ra \chi_1^0 \chi_4^0$ or $\chi_2^0 \chi_4^0$.
Again, this can be qualitatively explained, up to phase space suppression
factors,  by recalling the approximate formulae of eqs.(4.18--19), since the
situation is the same as for $H$, with the two neutralino states $\chi_3^0$ and
$\chi_4^0$ being interchanged. \s 

For small values of the common gaugino mass, $m_{1/2} \lsim 100$ GeV, the decay
mode of the pseudoscalar Higgs boson into stop squarks, $A\ra \tilde{t}_1
\bar{\tilde{t}}_2$, is phase space allowed. In this case, it is competitive
with the top--antitop decay mode. As discussed previously, the $1/M_A^2$
suppression [and to a lesser extent the suppression due to the mixing angle]
of the $A\ra \tilde{t}_1 \bar{\tilde{t}}_2$ decay width compared to 
$\Gamma (A \ra t\bar{t})$ will be compensated by the enhancement of the 
$A \tilde{t}_1 \bar{\tilde{t}}_2$ coupling for large values of $\mu$ and 
$A_t$. \s

Fig.8a shows the decay widths for the charged Higgs boson. Since the dominant
decay channel $H^+ \ra t\bar{b}$ is already open for $M_{H^\pm} \simeq 300$ 
GeV [although still slightly suppressed by phase space], the charged 
Higgs decay width into standard particles is rather large and it increases
only by a factor of $\sim 4$ when increasing the pseudoscalar mass to 
$M_A=600$ GeV. 
The situation for the supersymmetric decays is quite similar for the two 
masses: the chargino/neutralinos decay modes have branching ratios of the 
order of a few ten percent, while the branching ratios for the decays 
into sleptons, when kinematically allowed, do not exceed the level of a few 
percent, as expected. Only the decay $H^+ \ra \tilde{t}_1 \tilde{b}_1$, 
the only squark decay mode allowed by phase space [see Fig.3c] for relatively 
low values of $m_{1/2}$, is competitive with the $t\bar{b}$  decay mode. \s

The decay widths of the charged Higgs into the various combinations of
charginos and neutralinos are shown in Fig.8b for $M_{H^\pm} \sim 600$ GeV. 
The dominant channels are again decays into mixtures of gauginos and 
higgsinos, since $|\mu|$ is large. The pattern follows approximately the 
rules of eq.(4.22), modulo phase suppression. \s

As discussed in section 2, since the chargino, neutralino and sfermion
masses scale as $M_A$, the situation for even larger values of the
pseudoscalar Higgs boson mass, $M_A \sim 1$ TeV, will be qualitatively
similar to what has been discussed for $M_A \sim 600$ GeV. The only
exception is that there will be slightly more phase space available for
the supersymmetric decays to occur. 


\setcounter{equation}{0}
\renewcommand{\theequation}{5.\arabic{equation}}

\subsection*{5. Final Decay Products of the Higgs Bosons}

In this section, we will qualitatively describe the final decay products
of the produced Higgs bosons. Assuming that $M_A$ is large, $M_A \gsim
500$ GeV, the decays into standard particles [and more precisely, the
$t\bar{t}$ for the neutral and the $t\bar{b}$ decays for the charged
Higgs bosons] always have substantial branching ratios, even for the
value $\tb=1.75$ which will be chosen for the discussion. Therefore, to
investigate decays into SUSY particles in the main production processes,
$\ee \ra HA$ and $H^+H^-$, one has to look for final states where one of
the Higgs bosons decays into standard modes while the other Higgs boson
decays into charginos, neutralinos or stop squarks. As discussed
previously, the decays into the other squarks are disfavored by phase space,
while the branching ratios into sleptons are always small
and can be neglected. \s 

Let us first discuss the case where one of the Higgs bosons decays into
chargino and neutralino pairs, 
\beq
\ee & \ra & H \ A \ \ra [\,t\bar{t}\,]\,[\,\chi^+ \chi^-\,] \ \ 
\mbox{and} \ \ 
[\,t\bar{t}\,]\,[\,\chi^0 \chi^0\,] 
\ \ \non \\
\ee & \ra & H^+ H^- \ra [\,tb\,]\,[\,\chi^\pm \chi^0\,] 
\eeq
The lightest chargino $\chi_1^+$ and next--to--lightest neutralino 
$\chi_{2}^0$ decay into [possibly virtual] $W,Z$ 
and the lightest Higgs boson $h$, assuming that decays into sleptons and 
squarks are kinematically disfavored. In the limit of large $|\mu|$, the 
decay widths [in the decoupling limit] are proportional to 
\cite{R30}
\beq
\Gamma( \chi_1^+ \ra  \chi_1^0 W^+) & \sim & \sin^2 2\beta \\
\Gamma( \chi_2^0 \ra  \chi_1^0 Z )  & \sim & \cos^2 2\beta 
\left[ (M_2-M_1)/2\mu  \right]^2 \non \\
\Gamma( \chi_2^0 \ra  \chi_1^0 h ) & \sim & \sin^2 2 \beta 
\eeq
In most of the parameter space, the $W/Z/h$ are virtual [in addition to the 
three--body phase space factors, the decay widths are suppressed by powers 
of $M_2 M_Z/\mu^2$] except near the upper values of $m_{1 /2}$.
In the case of $\chi_2^0$, the channel $ \chi_2^0 \ra \chi_1^0 Z$ mode is 
always dominant although suppressed by additional powers of $M_2^2/\mu^2$ 
compared to the $ \chi_2^0 \ra h \chi_1^0$ mode, since both $h$ and $Z$ are 
off--shell, and the
$Z$ boson width is much larger than the width of the $h$ boson for small
values of $\tb$. The radiative decay $\chi_2^0 \ra \chi_1^0 \gamma$
should play a marginal role except for very small values of $m_{1 /
2}$ where the difference between the $\chi_2^0$ and $Z$ boson masses
becomes too large. \s 

For large values of $m_{1 /2}$, the sleptons become rather light 
compared to the gauginos
and the decays of the light chargino and neutralino into
leptons+sleptons are kinematically possible. In this case, these cascade
decays become dominant since the partial widths for large $|\mu|$ are
given by 
\beq
\sum_{l} \Gamma(\chi_2^0 \ra l \tilde{l}) =
\sum_{l} 2 \Gamma(\chi_1^\pm \ra l \tilde{\nu}) = \frac{3 G_F^2 M_W^2}{\sqrt{2}
\pi} M_2 
\eeq
and therefore not suppressed by powers of $M_Z M_2/\mu^2$, unlike the
previous decay modes [we assume of course that there is no suppression by
phase--space]. The sleptons will then decay into the LSP and massless
leptons, leading to multi--lepton final states. \s 

The heavier chargino, in the absence of squark and slepton decay modes, will
decay preferentially into the lightest chargino and neutralinos plus 
gauge or light Higgs bosons. The decay widths, in units of 
$G_FM_W^2 |\mu| /(8\sqrt{2}\pi)$ may be approximated in the 
decoupling limit by \cite{R30}
\beq
\chi_2^+ &  \ra &   \  \chi_1^+ Z \  \ : \ \Gamma = 1  \non \\
         & \ra &  \  \chi_1^+ h \  \ : \ \Gamma = 1 \non \\
         & \ra & \  \chi_1^0 W^+ \  \ : \ \Gamma = \tan ^2 \theta_W \non \\
         & \ra & \  \chi_2^0 W^+ \  \ : \ \Gamma = 1 
\eeq
The branching ratios for the various final states are roughly equal. Since 
$\chi_2^+$ is almost higgsino--like, the decay  widths into sleptons and 
partners of the light quarks, when kinematically allowed, are extremely 
small since they are suppressed by powers of $m_f^2/M_Z^2$. Because
of the large $m_t$ value, only the decays into stop squarks and bottom 
quarks will be very important. This decay is allowed in most of the parameter 
space for $M_A \gsim 600$ GeV and, up to suppression by 
mixing angles, it is enhanced by a power $m_t^2$ \cite{R30}
\beq
\frac{\Gamma( \chi_2^+ \ra \tilde{t} b )} 
{\Gamma( \chi_2^+ \ra W,Z,h) } \sim \frac{3m_t^2}{M_W^2} \frac{1}{\sin^2\beta
(3+\tan^2\theta_W)} \sim 4
\eeq
compared to the other decays. Therefore, when kinematically possible, 
this decay will be the dominant decay mode of the heavy charginos. \s

For the heavier neutralinos, $\chi_{3,4}^0$, the decay
widths into $W/Z/h$ bosons, again in units of $G_FM_W^2|\mu| 
/(8\sqrt{2}\pi)$ may be be written in the decoupling limit as \cite{R30}
\beq
\chi_{3/4}^0 & \ra & \  \chi_1^0 Z \ \  \ : \ \Gamma = \frac{1}{2}
\tan^2 \theta_W (1\pm \sin 2 \beta)   \non \\
             & \ra & \  \chi_1^0 h \ \  \ : \ \Gamma = \frac{1}{2}
\tan^2 \theta_W (1\mp \sin 2 \beta)   \non \\
             & \ra &  \  \chi_2^0 Z \ \  \ : \ \Gamma = \frac{1}{2} (1\pm 
\sin 2 \beta)   \non \\
             & \ra & \ \chi_2^0 h \ \  \ : \ \Gamma = \frac{1}{2} (1 \mp 
\sin 2 \beta)   \non \\
             &  \ra & \  \chi_1^+ W^- \ \  \ : \ \Gamma = 2 
\eeq
The dominant mode is the charged decay, $\chi^0_{3,4} \ra \chi_1^+W^- $,
followed by the modes involving the $h(Z)$ boson for $\chi^0_4
(\chi^0_3)$. Because $\sin 2\beta \sim 1$, only one of the $h$ or $Z$
decay channels is important. Here again, because of the higgsino nature
of the two heavy neutralinos, the decay widths into sleptons and the
scalar partners of the light quarks are negligible; the only important
decays are the stop decays, $\chi_{3,4}^0 \ra t \tilde{t}_1$,  when they
are allowed kinematically [i.e. for not too large values of $m_{1 /
2}]$. The ratio between stop and $W/Z/h$ decay widths, up to suppression 
by mixing angles, is also given by eq.(5.6), and the stop decays will 
therefore dominate. 

\bigskip

We now turn to the case where one of the produced Higgs particles decays 
into stop squarks
\beq
\ee & \ra & \ H \ A \ \ra [\,t\bar{t}\,]\,[\,\tilde{t}_1 \tilde{t}_1\,] \ \  
\mbox{and} \ \ [\,t\bar{t}\,]\,[\,\tilde{t}_1 \tilde{t_2}\,] 
\ \ \non \\
\ee & \ra & H^+ H^- \ra [tb][\tilde{t}_1 \tilde{b}_1] 
\eeq
{}From the squark mass plots, Fig.~3c, the only decay modes of the lightest 
stop squark allowed by phase space are
\beq
\tilde{t}_1 \ra t\chi_1^0 \ \ , \ \ t\chi_2^0 \ \ , \ \ b \chi_1^+
\eeq
Only the last decay mode occurs for relatively small values of $m_{1 
/2}$, since $m_{\tilde{t}_1} < m_t + m_{\chi^0_{1,2}}$ in this case. 
For larger values of 
$m_{1 /2}$, $\tilde{t}_1$ is heavy enough to decay into top quarks plus the 
lightest neutralinos. For these $m_{1 /2}$ values, the three decay
modes of eq.(5.9) will have approximately the same magnitude since the chargino 
and the neutralinos are gaugino--like and there is no enhancement due to 
the top mass for the $\tilde{t}_1 \ra t \chi^0$ decays. \s

The heavier stop squark, in addition to the previous modes, has 
decay channels with $\tilde{t}_1$ and $Z/h$ bosons in the
final state 
\beq
\tilde{t}_2 \ra \tilde{t}_1 Z \ \ , \ \ \tilde{t}_1 h 
\eeq
These decays, in particular the decay into the lightest Higgs boson
$h$, will be dominant in the large $|\mu|$ limit, since they will be 
enhanced by powers of $\mu^2$.

\newpage

\setcounter{equation}{0}
\renewcommand{\theequation}{A\arabic{equation}}

\subsection*{Appendix A: Chargino and Neutralino Masses and Couplings}

In this Appendix we collect the analytical expressions of the chargino 
and neutralino masses and couplings, and we discuss the limit 
in which the  Higgs--higgsino mass parameter $|\mu|$ is large. \s

The general chargino mass matrix \cite{R2a},
\begin{eqnarray}
{\cal M}_C = \left[ \begin{array}{cc} M_2 & \sqrt{2}M_W \sin \beta
\\ \sqrt{2}M_W \cos \beta & \mu \end{array} \right]
\end{eqnarray}
is diagonalized by two real matrices $U$ and $V$, 
\begin{eqnarray}
U^* {\cal M}_C V^{-1} \ \ \ra \ \ U={\cal O}_- \ {\rm and} \ \ V = 
\left\{
\begin{array}{cc} {\cal O}_+ \ \ \ & {\rm if \ det}{\cal M}_C >0  \\
            \sigma  {\cal O}_+ \ \ \ & {\rm if \ det}{\cal M}_C <0  
\end{array}
\right. 
\end{eqnarray}
where $\sigma$ is the matrix 
\begin{eqnarray}
{\sigma} = \left[ \begin{array}{cc} \pm 1 & 0
\\ 0 & \pm 1 \end{array} \right] 
\end{eqnarray}
with the appropriate signs depending upon the values of $M_2$, $\mu$, and
$\tan\beta$ in the chargino mass matrix.
${\cal O}_\pm$ is given by:
\begin{eqnarray}
{\cal O}_\pm = \left[ \begin{array}{cc} \cos \theta_\pm & \sin \theta_\pm
\\ -\sin \theta_\pm & \cos \theta_\pm \end{array} \right] 
\end{eqnarray}
with
\begin{eqnarray}
\tan 2 \theta_- &= & \frac{ 2\sqrt{2}M_W(M_2 \cos \beta
+\mu \sin \beta)}{ M_2^2-\mu^2-2M_W^2 \cos \beta} \non \\
\tan 2 \theta_+ & = & \frac{ 2\sqrt{2}M_W(M_2 \sin \beta
+\mu \cos \beta)}{M_2^2-\mu^2 +2M_W^2 \cos \beta} 
\end{eqnarray}
This leads to the two chargino masses, the 
$\chi_{1,2}^+$ masses
\begin{eqnarray}
m_{\chi_{1,2}^+} = && \frac{1}{\sqrt{2}} \left[ M_2^2+\mu^2+2M_W^2
\right. \non \\
&& \left. \mp \left\{ (M_2^2-\mu^2)^2+4 M_W^4 \cos^2 2\beta+4M_W^2 (M^2_2+\mu^2
+2M_2\mu \sin 2\beta) \right\}^{\frac{1}{2}} \right]^{\frac{1}{2}}
\end{eqnarray}
In the limit $|\mu| \gg M_2, M_Z$, the masses of the two charginos reduce to
\begin{eqnarray}
m_{\chi_{1}^+} & \simeq &  M_2 - \frac{M_W^2}{\mu^2} 
\left( M_2 +\mu \sin 2 \beta
\right) \non \\
m_{\chi_{2}^+} & \simeq & |\mu| + 
\frac{M_W^2}{\mu^2} \epsilon_\mu \left( M_2 \sin 
2 \beta +\mu \right) 
\end{eqnarray}
where $\epsilon_\mu$ is for the sign of $\mu$. For $|\mu| \ra \infty$,
the lightest chargino corresponds to a pure wino state with mass 
$m_{\chi_{1}^+} \simeq M_2$, while the heavier chargino corresponds to a 
pure higgsino state with a mass $m_{\chi_{1}^+} = |\mu|$.  

\bigskip 

In the case of the neutralinos, the four-dimensional neutralino mass matrix 
depends on the same two mass parameters $\mu$ and $M_2$, if the GUT 
relation
$M_1=\frac{5}{3} \tan^2 \theta_W$ $ M_2 \simeq \frac{1}{2} M_2$ 
\cite{R2a} is used. In the $(-i\tilde{B}, -i\tilde{W}_3, \tilde{H}^0_1,$ 
$\tilde{H}^0_2)$ basis, it has the form  
\begin{eqnarray}
{\cal M}_N = \left[ \begin{array}{cccc}
M_1 & 0 & -M_Z s_W \cos\beta & M_Z  s_W \sin\beta \\
0   & M_2 & M_Z c_W \cos\beta & -M_Z  c_W \sin\beta \\
-M_Z s_W \cos\beta & M_Z  c_W \cos\beta & 0 & -\mu \\
M_Z s_W \sin \beta & -M_Z  c_W \sin\beta & -\mu & 0
\end{array} \right]
\end{eqnarray}

\smallskip

It can be diagonalized analytically \cite{R31} by a single real matrix $Z$;
the [positive] masses of the neutralino states $m_{\chi_i^0}$ are given by
\beq
\epsilon_1 m_{\chi_1^0} &=& C_1 -\left( \frac{1}{2} a- \frac{1}{6}C_2
\right)^{1/2} + \left[ - \frac{1}{2} a- \frac{1}{3}C_2 + \frac{C_3}
{(8a-8C_2/3)^{1/2}} \right]^{1/2} \non \\
\epsilon_2 m_{\chi_2^0} &=& C_1 +\left( \frac{1}{2} a- \frac{1}{6}C_2
\right)^{1/2} - \left[ - \frac{1}{2} a- \frac{1}{3}C_2 - \frac{C_3}
{(8a-8C_2/3)^{1/2}} \right]^{1/2} \non \\
\epsilon_3 m_{\chi_3^0} &=& C_1 -\left( \frac{1}{2} a- \frac{1}{6}C_2
\right)^{1/2} - \left[ - \frac{1}{2} a- \frac{1}{3}C_2 + \frac{C_3} 
{(8a-8C_2/3)^{1/2}} \right]^{1/2} \non \\
\epsilon_4 m_{\chi_4^0} &=& C_1 +\left( \frac{1}{2} a- \frac{1}{6}C_2
\right)^{1/2} + \left[ - \frac{1}{2} a- \frac{1}{3}C_2 - \frac{C_3}
{(8a-8C_2/3)^{1/2}} \right]^{1/2}
\eeq
where $\epsilon_i = \pm 1$; the coefficients $C_i$ and $a$ are given by
\beq
C_1 &=& (M_1+M_2)/4 \non \\
C_2 &=& M_1 M_2 - M_Z^2 -\mu^2 -6 C_1^ 2 \non \\
C_3 &=& 2 C_1 \left[ C_2 + 2 C_1^2 +2 \mu^2 \right]+
M_Z^2 (M_1 c_W^2 + M_2 s_W^2) - \mu M_Z^2 \sin 2 \beta \non \\
C_4 &=& C_1 C_3- C_1^2 C_2 -C_1^4 -M_1 M_2 \mu^2 +(M_1 c_W^2 + M_2 s_W^2)
M_Z^2 \mu \sin 2\beta
\eeq
and
\beq
a = \frac{1} {2^{1/3}} {\rm Re} \left[ S+ i \left( \frac{D}{27} 
\right)^{1/2} \right]^{1/3}
\eeq
with
\beq
S &=& C_3^2+\frac{2}{27} C_2^3 -\frac{8}{3} C_2 C_4 \non \\
D &=& \frac{4}{27} (C_2^2 +12 C_4)^3 -27 S^2 
\eeq

In the limit of large $|\mu|$ values, the masses of the neutralino states 
simplify to 
\begin{eqnarray}
m_{\chi_{1}^0} &\simeq& M_1 - \frac{M_Z^2}{\mu^2} \left( M_1 +\mu \sin 2 \beta
\right) s_W^2 \non \\
m_{\chi_{2}^0} &\simeq& M_2 - \frac{M_Z^2}{\mu^2} \left( M_2 +\mu \sin 2 \beta
\right) c_W^2 \non \\
m_{\chi_{3}^0} &\simeq& |\mu| + \frac{1}{2}\frac{M_Z^2}{\mu^2} \epsilon_\mu 
(1-\sin 2\beta) \left( \mu + M_2 s_W^2+M_1 c_W^2 \right) \non \\
m_{\chi_{4}^0} &\simeq& |\mu| + \frac{1}{2}\frac{M_Z^2}{\mu^2} \epsilon_\mu 
(1+\sin 2\beta) \left( \mu - M_2 s_W^2 - M_1 c_W^2 \right) 
\end{eqnarray}
Again, for $|\mu| \ra \infty$, two neutralinos are pure gaugino states 
with masses $m_{\chi_{1}^0} \simeq M_1$ , $m_{\chi_{2}^0} =M_2$, while
the two others are pure higgsino states, with masses 
$m_{\chi_{3}^0} \simeq m_{\chi_{4}^0} \simeq |\mu|$. 

\bigskip

The matrix elements of the diagonalizing matrix, $Z_{ij}$ with $i,j=1,..4$,
are given by
\begin{eqnarray}
Z_{i1} &=& \left[ 1+ \left(\frac{Z_{i2}}{Z_{i1}}\right)^2+\left(\frac{Z_{i3}}
{Z_{i1}}\right)^2+\left(\frac{Z_{i4}}{Z_{i1}}\right)^2 \right]^{-1/2} \\
\frac{Z_{i2}}{Z_{i1}} &=&  -\frac{1}{\tan \theta_W} \frac{ M_1-
\epsilon_i  m_{\chi_{i}^0} } {M_2 -\epsilon_i m_{\chi_{i}^0} } \non \\
\frac{Z_{i3}}{Z_{i1}} &=&  \frac{
\mu (M_1-\epsilon_i  m_{\chi_{i}^0} )(M_2 -\epsilon_i m_{\chi_{i}^0} )
-M_Z^2 \sin \beta \cos \beta [(M_1-M_2)c_W^2+M_2 -\epsilon_i m_{\chi_{i}^0}] } 
{M_Z (M_2 -\epsilon_i m_{\chi_{i}^0} )s_W [\mu\cos \beta 
+\epsilon_i m_{\chi_{i}^0} \sin \beta) } \non \\
\frac{Z_{i4}}{Z_{i1}} &=&  \frac{
-\epsilon_i m_{\chi_{i}^0} (M_1-\epsilon_i m_{\chi_{i}^0} )(M_2 -\epsilon_i 
m_{\chi_{i}^0} ) -M_Z^2 \cos^2 \beta [(M_1-M_2)c_W^2+M_2 
-\epsilon_i  m_{\chi_{i}^0}] } 
{M_Z (M_2 -\epsilon_i m_{\chi_{i}^0} )s_W [\mu\cos \beta
+\epsilon_i m_{\chi_{i}^0} \sin \beta) } \non 
\end{eqnarray}
where $\epsilon_i$ is the sign of the $i$th eigenvalue of the neutralino 
mass matrix, which in the large $|\mu|$ limit are: $\epsilon_1
=\epsilon_2=1$ and $\epsilon_4=-\epsilon_3=\epsilon_\mu$. 

\newpage

\setcounter{equation}{0}
\renewcommand{\theequation}{B\arabic{equation}}

\subsection*{Appendix B: Sfermion Masses and Mixing}

We now present the explicit expressions of the squark and slepton masses. 
We will assume a universal scalar mass $m_0$ and gaugino mass $m_{1/2}$ at 
the GUT scale, and we will neglect the Yukawa couplings in the RGE's 
[see Appendix C]. For third generation squarks this is a poor approximation 
since these couplings can be large; these have been taken into account in 
the numerical analysis. \s

By performing the RGE evolution to the electroweak scale, one obtains for the
left-- and right--handed sfermion masses at one--loop order [we include the 
full two--loop evolution of the masses in the numerical analysis] 
\begin{eqnarray}
m_{\tilde{f}_{L,R}}^2 = m_0^2 + \sum_{i=1}^{3} F_i (f) 
m^2_{1/ 2} \pm ( T_{3 {f} } - e_{f} s_W^2 ) M_Z^2 
\cos 2 \beta
\end{eqnarray}
$T_{ 3 {f} }$ and $e_{{f}}$ are the weak isospin and the 
electric charge of the corresponding fermion ${f}$, and $F_i$ are the RGE 
coefficients for the three gauge couplings at 
the scale $Q \sim M_Z$, given by
\begin{eqnarray}
F_i = \frac{c_i(f)}{b_i} \left[1- \left( 1 - \frac{\alpha_G}{4\pi} 
b_i \log \frac{Q^2}{M_G^2} \right)^{-2} \right] 
\end{eqnarray}
The coefficients $b_i$, assuming that all the MSSM particle spectrum 
contributes to the evolution from $Q$ to the GUT scale $M_G$, are
\begin{eqnarray}
b_1=33/5 \ \  , \ \ b_2=1 \ \ , \ \ b_3=-3
\end{eqnarray}
The coefficients $c(\tilde{f})=(c_1,c_2,c_3) (\tilde{f})$ depend on the 
hypercharge and color of the sfermions  [$F_L=L_L$ or $Q_L$ is the 
slepton or squark doublet]
\begin{eqnarray}
c(\tilde{L}_L) = \left( \begin{array}{c} 3/ 10 \\ 3/2\\ 0 \end{array} \right) 
\ , \ 
c(\tilde{E}_R)= \left( \begin{array}{c} 6/5 \\ 0 \\ 0 
\end{array} \right) \non 
\end{eqnarray}
\begin{eqnarray}
c(\tilde{Q}_L)= \left( \begin{array}{c} 1/30 \\ 3/2 \\ 
8/3 \end{array} \right) \ , \ 
c(\tilde{U}_R)= \left( \begin{array}{c} 8/15 \\ 0 \\ 8/3 
\end{array} \right) \ , \ 
c(\tilde{D}_R)= \left( \begin{array}{c} 2/15 \\ 0 \\ 8/3
 \end{array} \right) 
\end{eqnarray}

With the input gauge coupling constants at the scale of the $Z$ boson mass
\beq
\alpha_1 (M_Z) \simeq 0.01 \ \ , \ \ 
\alpha_2 (M_Z) \simeq 0.033 \ \ ,  \ \ 
\alpha_3 (M_Z) \simeq 0.118 
\eeq
one obtains for the GUT scale $M_G$ and for the coupling constant $\alpha_G$
\beq
M_G \sim 1.9 \times 10^{16} \ {\rm GeV} \ \ \ {\rm and} \ \ 
\alpha_G = 0.041
\eeq
Using these values, and including only gauge loops in the one--loop 
RGE's, one obtains for the left-- and right--handed sfermion masses
\cite{DN}
\beq
m^2_{\tilde{u}_L} &=& m_0^2 +6.28 m^2_{1/2} +0.35 M_Z^2 \cos(2\beta) \non \\
m^2_{\tilde{d}_L} &=& m_0^2 +6.28 m^2_{1/2} -0.42 M_Z^2 \cos(2\beta) \non \\
m^2_{\tilde{u}_R} &=& m_0^2 +5.87 m^2_{1/2} +0.16 M_Z^2 \cos(2\beta) \non \\
m^2_{\tilde{d}_R} &=& m_0^2 +5.82 m^2_{1/2} -0.08 M_Z^2 \cos(2\beta) \non \\
m^2_{\tilde{\nu}_L} &=& m_0^2 +0.52 m^2_{1/2} +0.50 M_Z^2 \cos(2\beta) \non \\
m^2_{\tilde{e}_L} &=& m_0^2 +0.52 m^2_{1/2} -0.27 M_Z^2 \cos(2\beta) \non \\
m^2_{\tilde{e}_R} &=& m_0^2 +0.15 m^2_{1/2} -0.23 M_Z^2 \cos(2\beta) 
\eeq

In the case of the third generation sparticles, left-- and right--handed 
sfermions will mix; for a given sfermion $\tilde{f} = \tilde{t}, \tilde{b}$
and $\tilde{\tau }$, the mass matrices which determine the mixing are 
\begin{eqnarray}
\left[ \begin{array}{cc} m_{\tilde{f}_L}^2 + m_f^2 & m_f (A_f - \mu r_f) 
\\ m_f (A_f - \mu r_f)  & m_{\tilde{f}_R}^2 + m_f^2 \end{array} \right]
\end{eqnarray}
where the sfermion masses $m_{\tilde{f}_{L,R}}$ are given above, $m_f$
are the masses of the partner fermions and $r_{b} = r_\tau =1/r_t= \tb$. 
These matrices are diagonalized by orthogonal matrices with mixing angles
$\theta_f$ defined by
\begin{eqnarray}
\sin 2\theta_f = \frac{2 m_f (A_f -\mu r_f)} { m_{\tilde{f}_1}^2
-m_{\tilde{f}_2}^2 } \ \ , \ \ 
\cos 2\theta_f = \frac{m_{\tilde{f}_L}^2 -m_{\tilde{f}_R}^2}
{ m_{\tilde{f}_1}^2 -m_{\tilde{f}_2}^2 } 
\end{eqnarray}
and the masses of the squark eigenstates given by
\begin{eqnarray}
m_{\tilde{f}_{1,2}}^2 = m_f^2 + \frac{1}{2} \left[ 
m_{ \tilde{f}_L}^2 + m_{\tilde{f}_R}^2 \mp \sqrt{
(m_{\tilde{f}_L}^2 - m_{\tilde{f}_R}^2)^2 + 4m_f^2 (A_f -\mu r_f)^2 } 
\right].
\end{eqnarray}

\newpage

\setcounter{equation}{0}
\renewcommand{\theequation}{C\arabic{equation}}

\subsection*{Appendix C: Renormalization Group Equations}

Finally, we collect for completeness the renormalization group equations for 
the soft--SUSY breaking parameters [the trilinear couplings, scalar masses 
as well as for $\mu$ and $B$], including the dependence on $A_t, A_b$ and 
$A_\tau$. We restrict ourselves to the one--loop RGE's and we keep only the leading 
terms in the mass hierarchy in the MSSM with three fermion generations. 
The complete expressions for the RGE's can be found 
in Refs.\cite{R15,R16}. \s

For the trilinear couplings of the third generation sfermions, the RGE's
are given by
\begin{eqnarray}
{{dA_t}\over {dt}}&=&{2\over {16\pi ^2}}\Big (\sum c_ig_i^2M_i
+6\lambda _t^2A_t+\lambda _b^2A_b\Big )\; \non \\
{{dA_b}\over {dt}}&=&{2\over {16\pi ^2}}\Big (\sum c_i^{\prime }g_i^2M_i
+6\lambda _b^2A_b+\lambda _t^2A_t+\lambda _{\tau}^2A_{\tau}\Big )\; \non \\
{{dA_{\tau }}\over {dt}}&=&{2\over {16\pi ^2}}\Big (\sum c_i^{\prime \prime }
g_i^2M_i+3\lambda _b^2A_b+4\lambda _{\tau}^2A_{\tau}\Big )\; 
\end{eqnarray}
while for the scalar masses of the third generation sfermions, one has
\begin{eqnarray}
{{dM_{Q_L}^2}\over {dt}}&=&{2 \over {16\pi ^2}}
\Big (-{1\over 15}g_1^2M_1^2-3g_2^2M_2^2-{16\over 3}g_3^2M_3^2
+\lambda _t^2X_t+\lambda _b^2X_b\Big )\; \non \\
{{dM_{t_R}^2}\over {dt}}&=&{2 \over {16\pi ^2}}
\Big (-{16\over 15}g_1^2M_1^2-{16\over 3}g_3^2M_3^2
+2\lambda _t^2X_t\Big )\; \non \\
{{dM_{b_R}^2}\over {dt}}&=&{2 \over {16\pi ^2}}
\Big (-{4\over 15}g_1^2M_1^2-{16\over 3}g_3^2M_3^2
+2\lambda _b^2X_b\Big )\; \non \\
{{dM_{L_L}^2}\over {dt}}&=&{2 \over {16\pi ^2}}
\Big (-{3\over 5}g_1^2M_1^2-3g_2^2M_2^2
+\lambda _{\tau}^2X_{\tau }\Big )\; \non \\
{{dM_{\tau _R}^2}\over {dt}}&=&{2 \over {16\pi ^2}}
\Big (-{12\over 5}g_1^2M_1^2
+2\lambda _{\tau}^2X_{\tau }\Big )\;
\end{eqnarray}
The evolution parameter is defined by $t=\log(Q/M_G)$,  
\begin{eqnarray}
b_i&=& (\, 33/5 \, , \, 1 \, , \, -3 \, ) \; \non \\
c_i&=&(\, 13/ 15 \, , \, 3 \, , \, 16/ 3 \, ) \; \non \\
c_i^{\prime}&=&(\, 7/ 15 \, , \, 3 \, , \, 16/ 3 \, ) \; \non \\
c_i^{\prime \prime}&=&(\, 9/ 5 \, , \, 3 \, , \, 0 \, ) \;
\end{eqnarray}
and
\begin{eqnarray} 
X_t      &=&  M_{Q_L}^2+M_{t _R}^2+M_{H_2}^2+A_t^2\; \non \\
X_b      &=&  M_{Q_L}^2+M_{b _R}^2+M_{H_1}^2+A_b^2\; \non \\
X_{\tau }&=&  M_{L_L}^2+M_{\tau _R}^2+M_{H_1}^2+A_{\tau }^2\;
\end{eqnarray}
For the first and second generation sfermions, these expressions 
reduce to
\begin{eqnarray}
{{dA_u}\over {dt}}&=&{2\over {16\pi ^2}}\Big (\sum c_ig_i^2M_i
+\lambda _t^2A_t\Big )\; \non \\
{{dA_d}\over {dt}}&=&{2\over {16\pi ^2}}\Big (\sum c_i^{\prime }g_i^2M_i
+\lambda _b^2A_b+{1\over 3}\lambda _{\tau}^2A_{\tau}\Big )\; \non \\
{{dA_e}\over {dt}}&=&{2\over {16\pi ^2}}\Big (\sum c_i^{\prime \prime }
g_i^2M_i+\lambda _b^2A_b+{1\over 3}\lambda _{\tau}^2A_{\tau}\Big )\;  
\end{eqnarray}
and
\begin{eqnarray}
{{dM_{q_L}^2}\over {dt}}&=&{2 \over {16\pi ^2}}
\Big (-{1\over 15}g_1^2M_1^2-3g_2^2M_2^2-{16\over 3}g_3^2M_3^2
\Big )\; \non \\
{{dM_{u_R}^2}\over {dt}}&=&{2 \over {16\pi ^2}}
\Big (-{16\over 15}g_1^2M_1^2-{16\over 3}g_3^2M_3^2
\Big )\; \non \\
{{dM_{d_R}^2}\over {dt}}&=&{2 \over {16\pi ^2}}
\Big (-{4\over 15}g_1^2M_1^2-{16\over 3}g_3^2M_3^2
\Big )\; \non \\
{{dM_{l_L}^2}\over {dt}}&=&{2 \over {16\pi ^2}}
\Big (-{3\over 5}g_1^2M_1^2-3g_2^2M_2^2
\Big )\; \non \\
{{dM_{e_R}^2}\over {dt}}&=&{2 \over {16\pi ^2}}
\Big (-{12\over 5}g_1^2M_1^2
\Big )\; 
\end{eqnarray}
For the gauge coupling constants and the other soft--SUSY breaking parameters, 
the RGE's are given by
\begin{eqnarray}
{{dg_i}\over {dt}}&=&{1\over {16\pi ^2}}b_ig_i^3 \;  \\
{{dM_i}\over {dt}}&=&{2\over {16\pi ^2}}b_ig_i^2  M_i\;  \\
{{dB}\over {dt}}&=&{2\over {16\pi ^2}}\Big ({3\over 5}g_1^2M_1+3g_2^2M_2
+3\lambda _b^2A_b+3\lambda _t^2A_t+\lambda _{\tau}^2A_{\tau}\Big )\; \\
{{d\mu }\over {dt}}&=&{\mu \over {16\pi ^2}}\Big (-{3\over 5}g_1^2-3g_2^2
+3\lambda _t^2+3\lambda _b^2+\lambda _{\tau}^2\Big )\; \\
{{dm_{H_1}^2}\over {dt}}&=&{2 \over {16\pi ^2}}
\Big (-{3\over 5}g_1^2M_1^2-3g_2^2M_2^2
+3\lambda _b^2X_b+\lambda _{\tau}^2X_{\tau }\Big )\; \\
{{dm_{H_2}^2}\over {dt}}&=&{2 \over {16\pi ^2}}
\Big (-{3\over 5}g_1^2M_1^2-3g_2^2M_2^2
+3\lambda _t^2X_t\Big )\;
\end{eqnarray}

\newpage

\begin{figure}[htbp]
\centerline{\psfig{figure=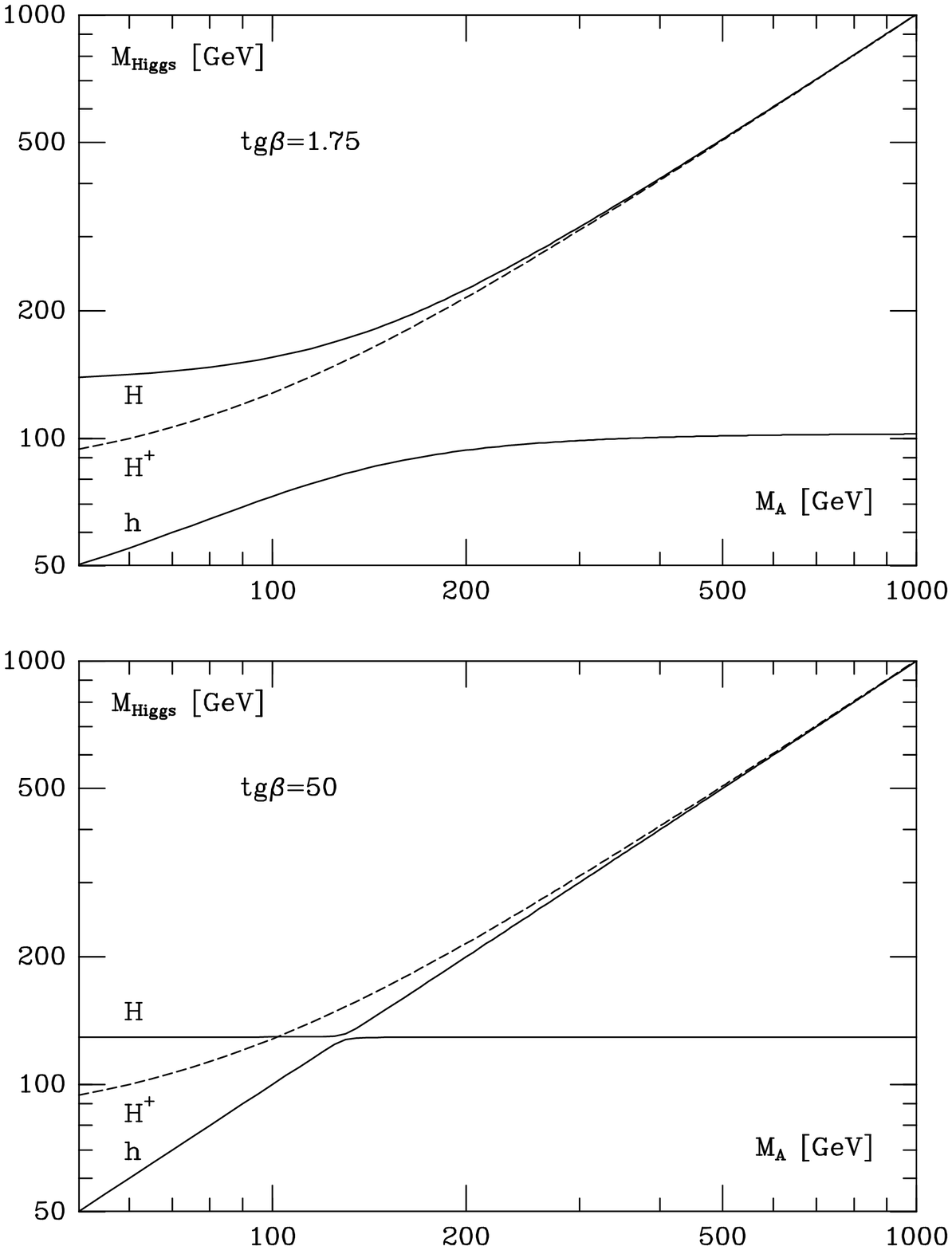,height=20.cm,width=16cm}}
\vspace*{-1.5cm}
\nn {\bf Fig.~1:}  Masses of the CP--even Higgs bosons $h,H$ and of the 
charged Higgs particles $H^\pm$ as a function of $M_A$ for two values of 
$\tb =1.75$ and 50; the common squark mass $M_S$ at the weak scale is fixed 
to $M_S =1$ TeV and we take $\mu=A_t=0$.
\end{figure}
\newpage

\begin{figure}[htbp]
\centerline{\psfig{figure=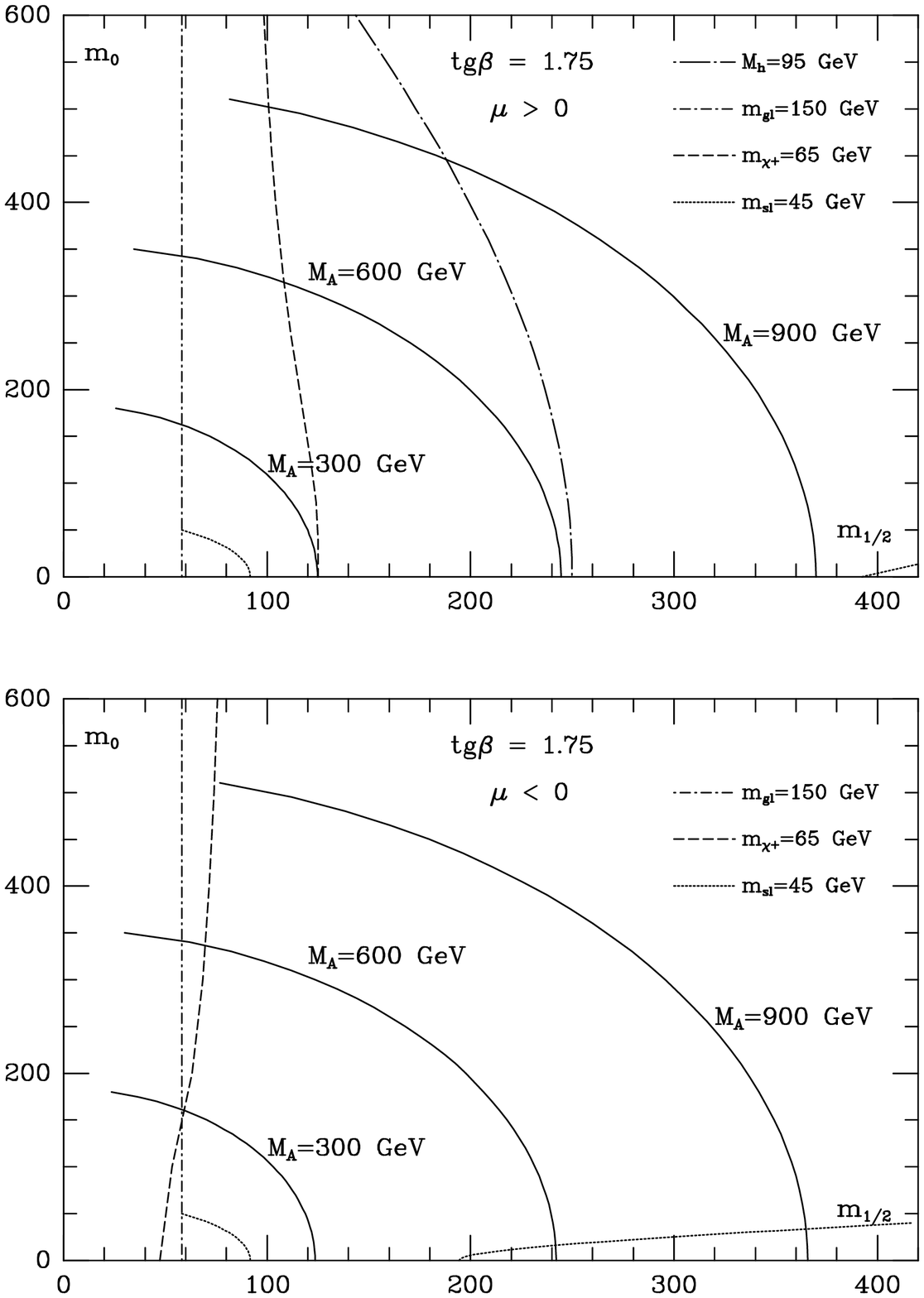,height=20.cm,width=16cm}}
\vspace*{-1.5cm}
\nn {\bf Fig.~2:}  The correlation between $m_0$ and $m_{1/2}$ for
$\tb=1.75$ and three values of $M_A=300, 600$ and 900 GeV. The non-solid 
lines show the boundaries which can be excluded by including the 
experimental bounds from LEP1.5 and Tevatron. 
\end{figure}
\newpage

\begin{figure}[htbp]
\centerline{\psfig{figure=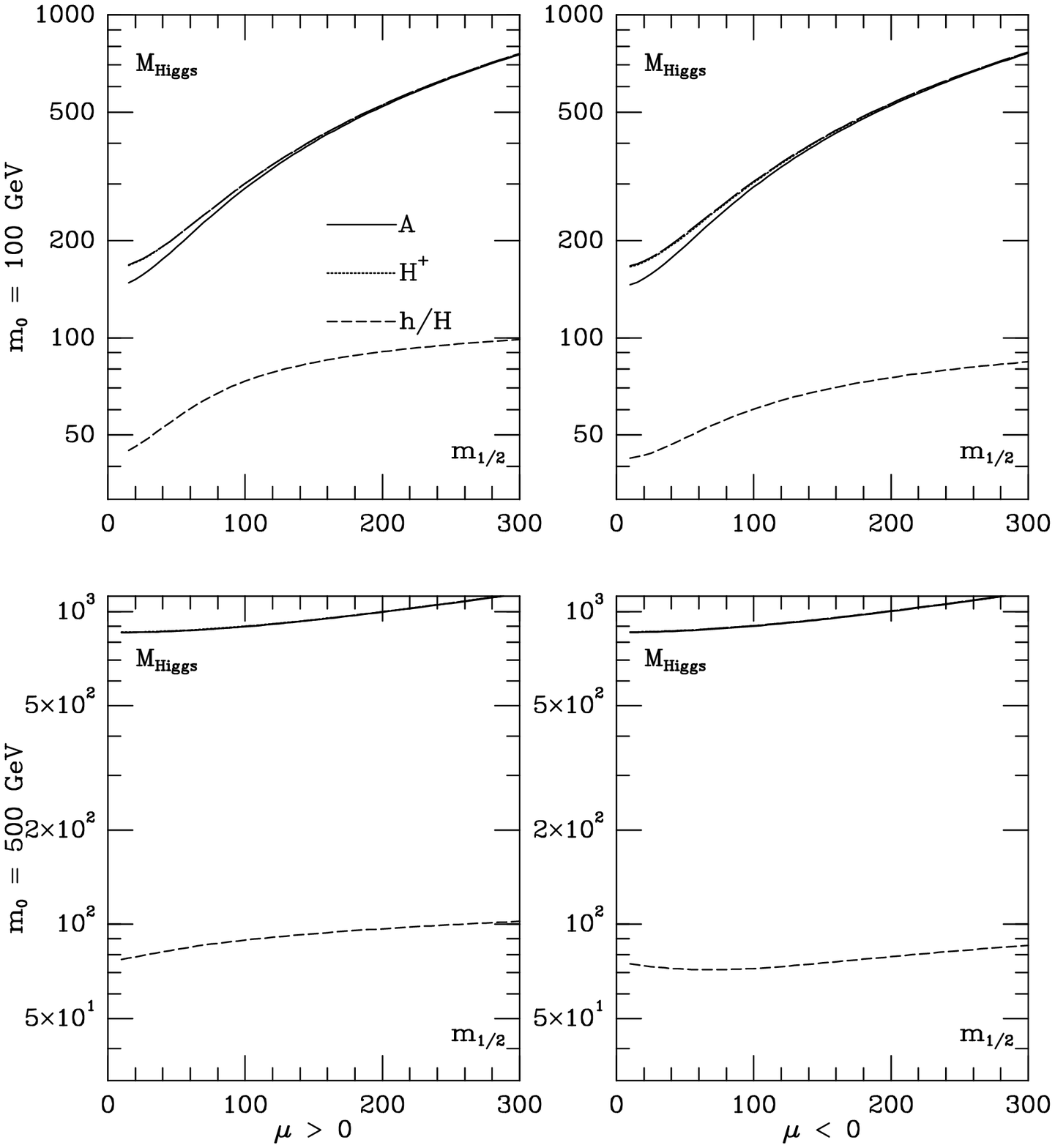,height=20.cm,width=16cm}}
\vspace*{-2.5cm}
\nn {\bf Fig.~3a:}  The masses of the Higgs bosons as a function of $m_{1/2}$ 
for $\tb =1.75$, for the two values $m_0=100$ and $500$ GeV and both
signs of $\mu$.
\end{figure}
\newpage

\begin{figure}[htbp]
\centerline{\psfig{figure=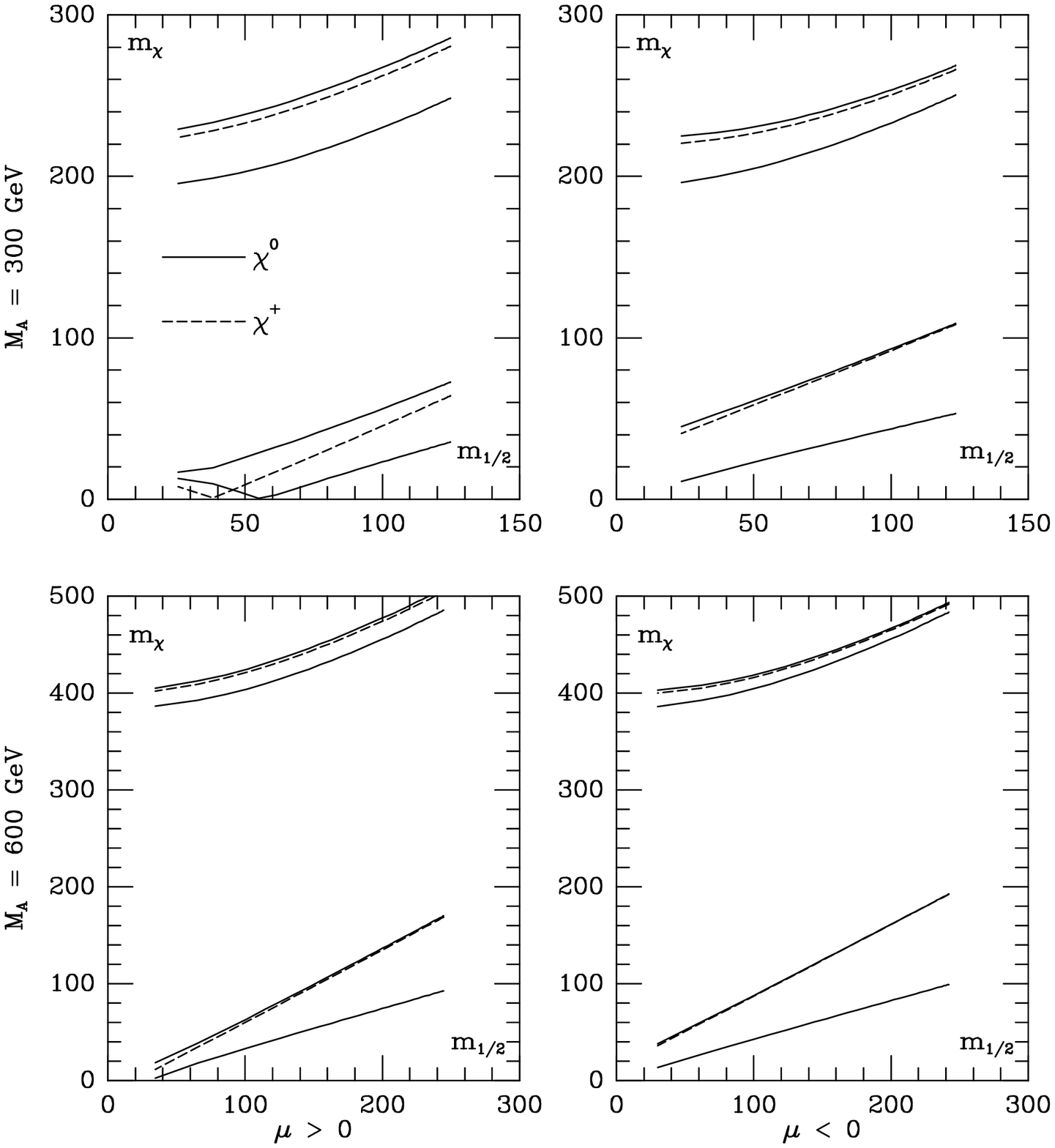,height=20.cm,width=16cm}}
\vspace*{-2.5cm}
\nn {\bf Fig.~3b:}  The masses of the two charginos (dashed lines) and the 
four neutralinos (solid lines) as a function of $m_{1/2}$ for $\tb 
=1.75$,  $M_A=300$ and $600$ GeV and for both signs of 
$\mu$. The chargins/neutralinos are ordered with increasing masses. 
\end{figure}
\newpage

\begin{figure}[htbp]
\centerline{\psfig{figure=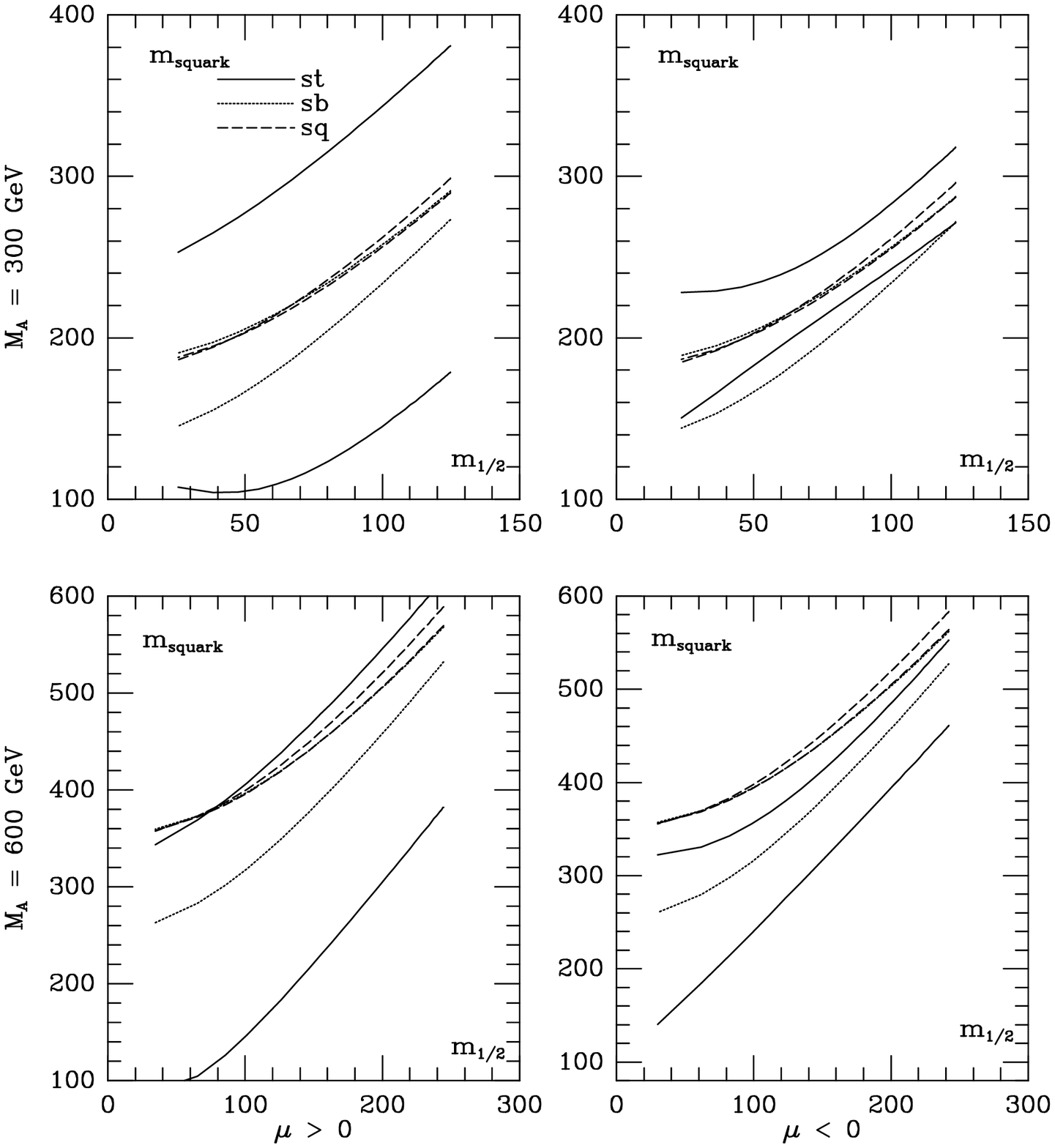,height=20.cm,width=16cm}}
\vspace*{-2.5cm}
\nn {\bf Fig.~3c:} The masses of the two stop (solid lines), sbottom 
(dotted lines) and first/second generation squark (dashed lines) 
eigenstates as a function of $m_{1/2}$ for $\tb =1.75$, 
$M_A=300$ and $600$ GeV and for both signs of $\mu$. 
\end{figure}
\newpage 

\begin{figure}[htbp]
\centerline{\psfig{figure=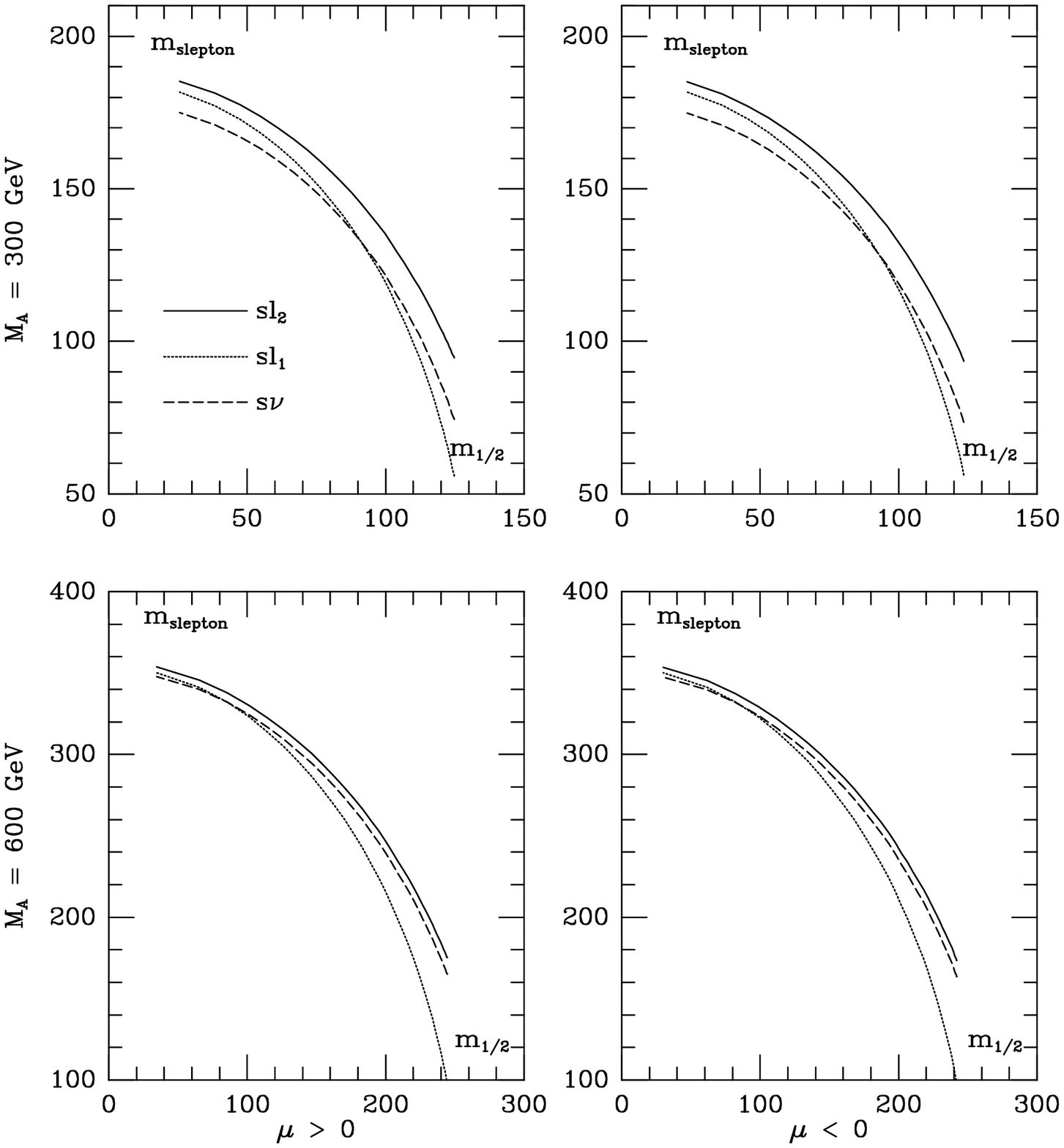,height=20.cm,width=16cm}}
\vspace*{-2.5cm}
\nn {\bf Fig.~3d:} The masses of the charged sleptons (solid and dotted
lines) and the sneutrino (dashed lines) of the three generations as a
function of $m_{1/2}$ for $\tb =1.75$,  $M_A=300$ and $600$
GeV and for both signs of $\mu$. 
\end{figure}
\newpage

\begin{figure}[htbp]
\centerline{\psfig{figure=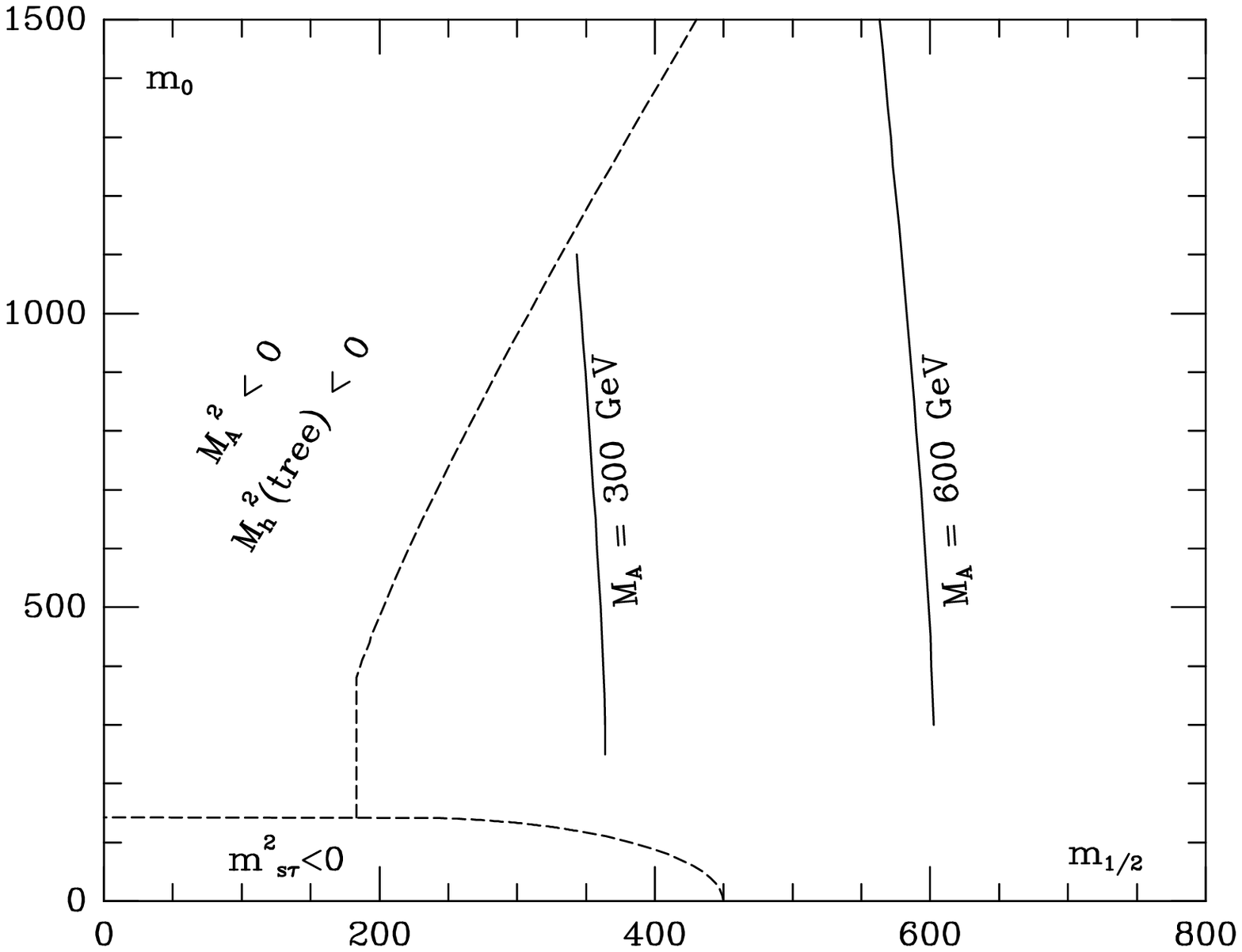,height=20.cm,width=16cm}}
\vspace*{-2.5cm}
\nn {\bf Fig.~4:}  The correlation between $m_0$ and $m_{1/2}$ for
$\tb \simeq 50$, $\mu <0$, and two values of $M_A=300$ and $600$ GeV. The boundary
contours correspond to tachyonic solutions, $m_{\tilde{\tau}}^2 <0$, $M_A^2<0$
and $M_h^2 <0$ at the tree--level. 
\end{figure}
\newpage

\begin{figure}[htbp]
\centerline{\psfig{figure=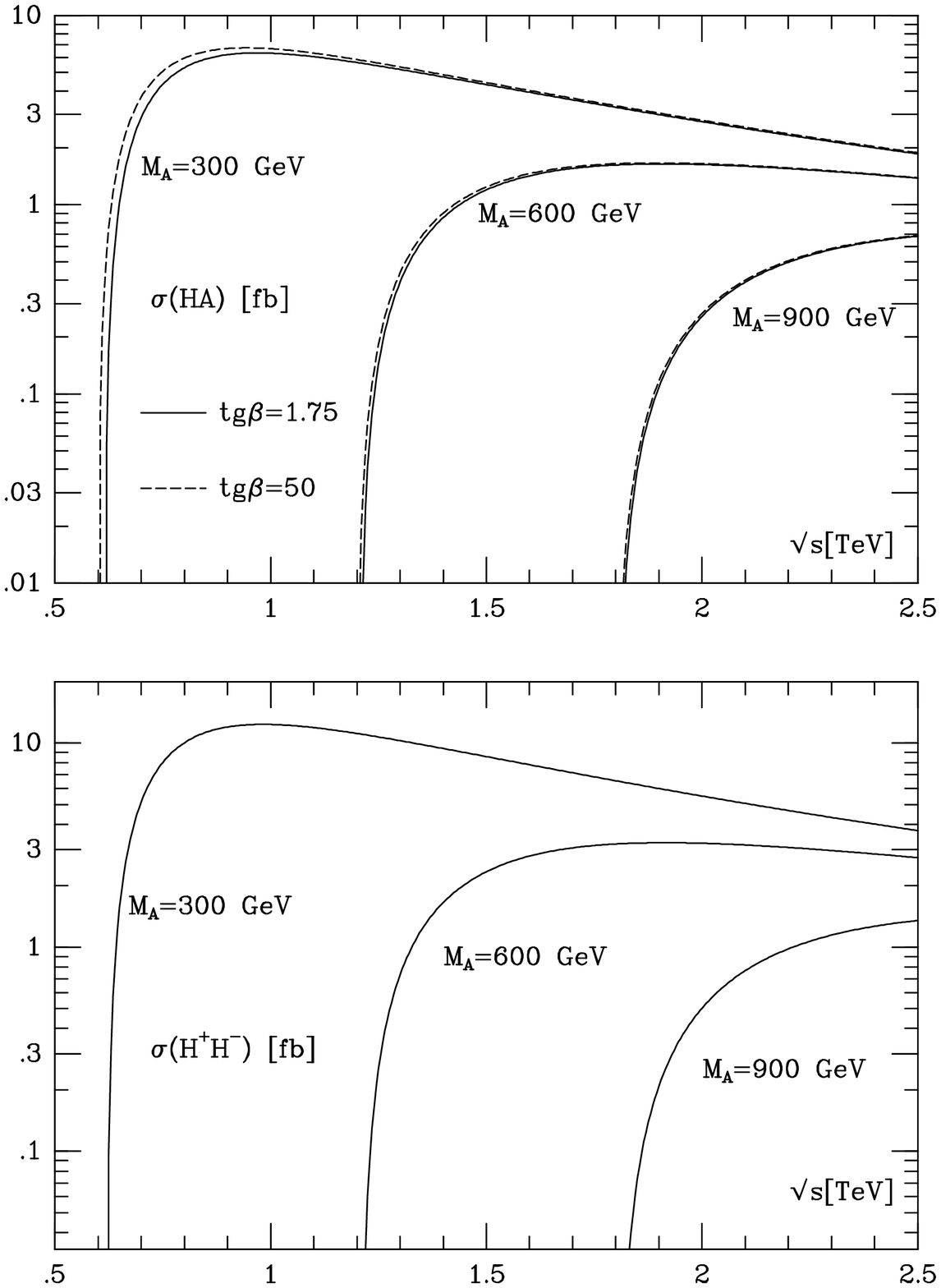,height=20.cm,width=16cm}}
\vspace*{-1.5cm}
\nn {\bf Fig.~5a:} Cross sections for the pair production processes
$\ee \ra HA$ and $\ee \ra H^+H^-$ as a function of $\sqrt{s}$ for $\tb=1.75$
(solid lines) and $\tb=50$ (dashed lines) and three values of $M_A=300,
600$ and $900$ GeV. 
\end{figure}
\newpage

\begin{figure}[htbp]
\centerline{\psfig{figure=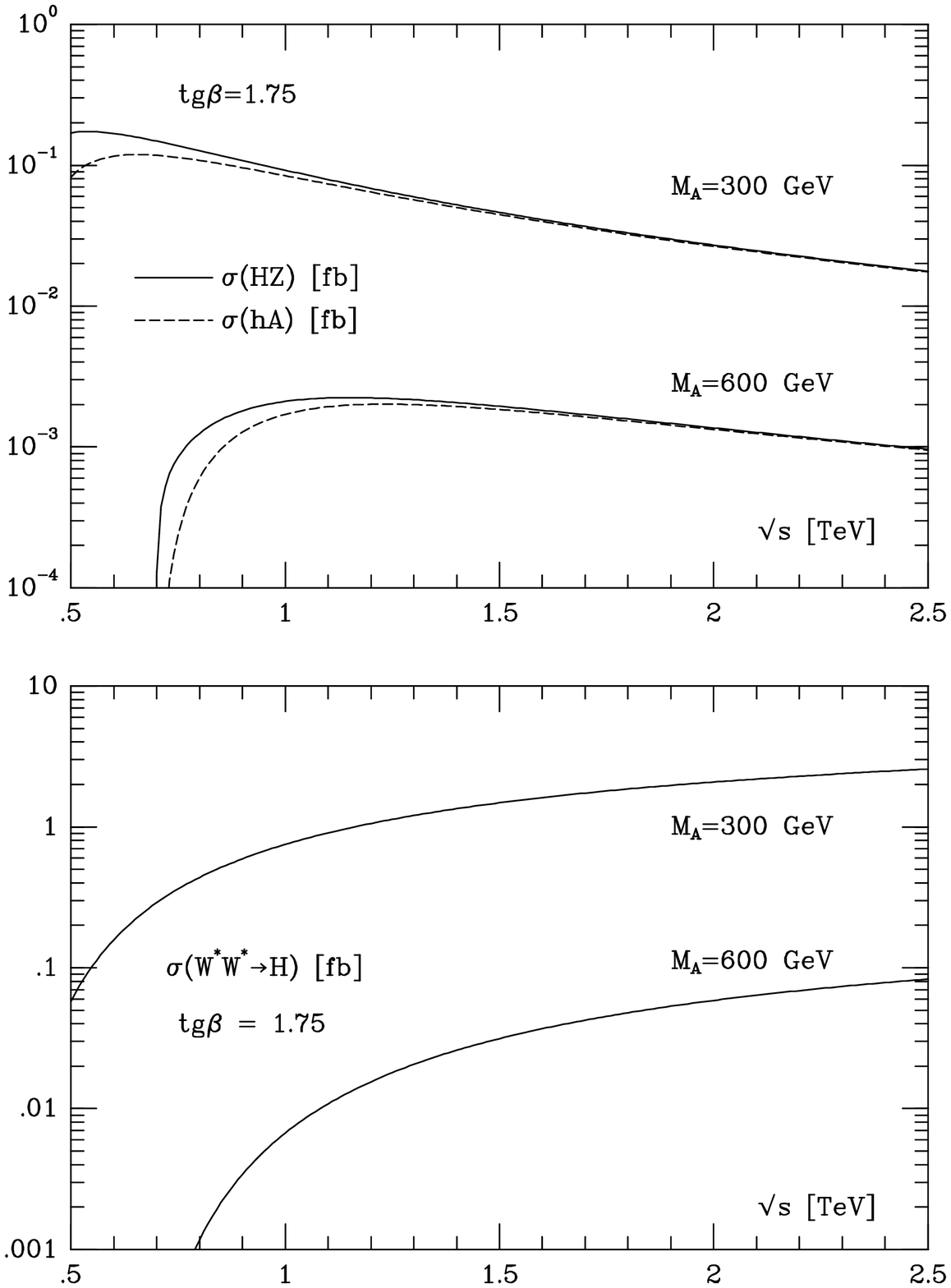,height=20.cm,width=16cm}}
\vspace*{-1.2cm}
\nn {\bf Fig.~5b:} Cross sections for the production processes $\ee \ra
HZ$, $\ee \ra hA$ and $\ee \ra H \nu \bar{\nu}$ as a function of 
$\sqrt{s}$ for $\tb=1.75$ and the values $M_A=300$ and $600$ GeV. 
\end{figure}
\newpage

\begin{figure}[htbp]
\centerline{\psfig{figure=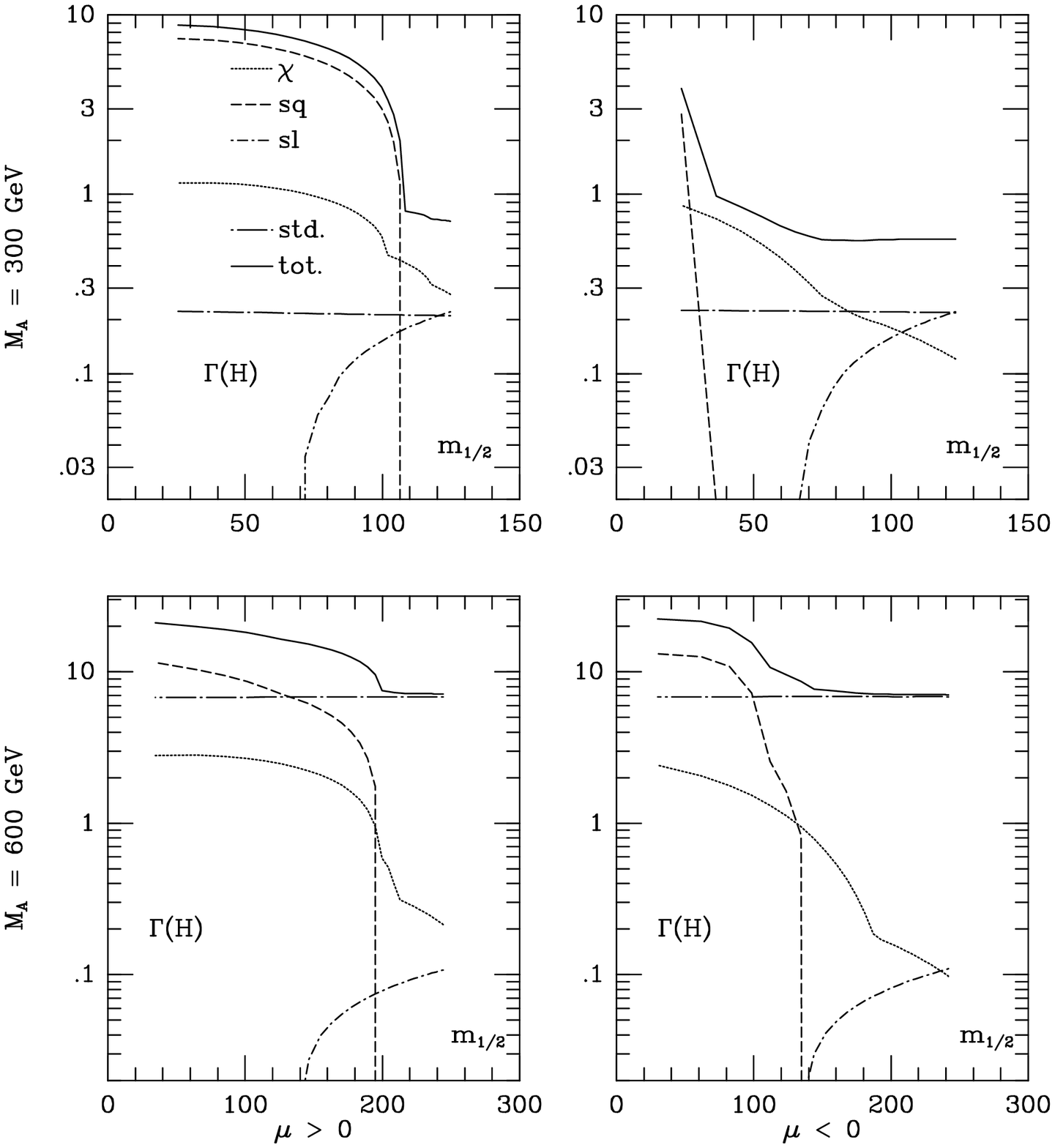,height=20.cm,width=16cm}}
\vspace*{-2.5cm}
\nn {\bf Fig.~6a:}  Decay widths (in GeV) of the heavy CP--even Higgs 
boson $H$ into charginos and neutralinos 
(dotted lines), squarks (dashed lines), 
sleptons (dash--dotted lines), standard particles (dott--long--dashed 
lines) 
and the total decay widths (solid lines) as a function of $m_{1/2}$ for 
$\tb =1.75$,  $M_A=300$ and $600$ GeV and for both signs of $\mu$. 
\end{figure}
\newpage

\begin{figure}[htbp]
\centerline{\psfig{figure=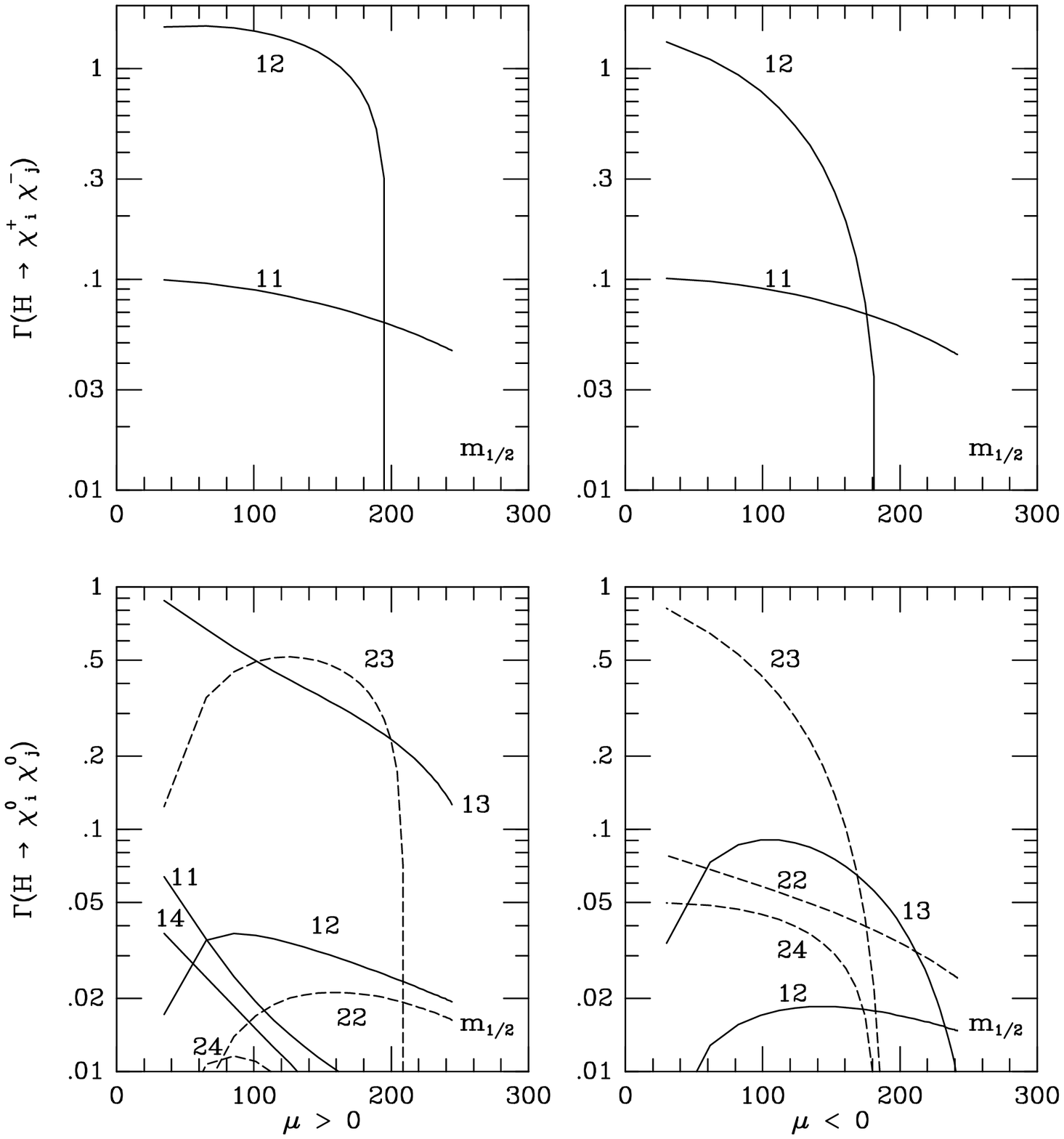,height=20.cm,width=16cm}}
\vspace*{-2.5cm}
\nn {\bf Fig.~6b:}  Partial decay widths (in GeV) of the heavy CP--even Higgs
boson $H$ into all combinations of chargino and neutralino pairs 
[$ij \equiv \chi_i \chi_j$]
as a function of $m_{1/2}$ for $\tb =1.75$, $M_A=600$ GeV 
and for both signs of $\mu$. 
\end{figure}
\newpage

\begin{figure}[htbp]
\centerline{\psfig{figure=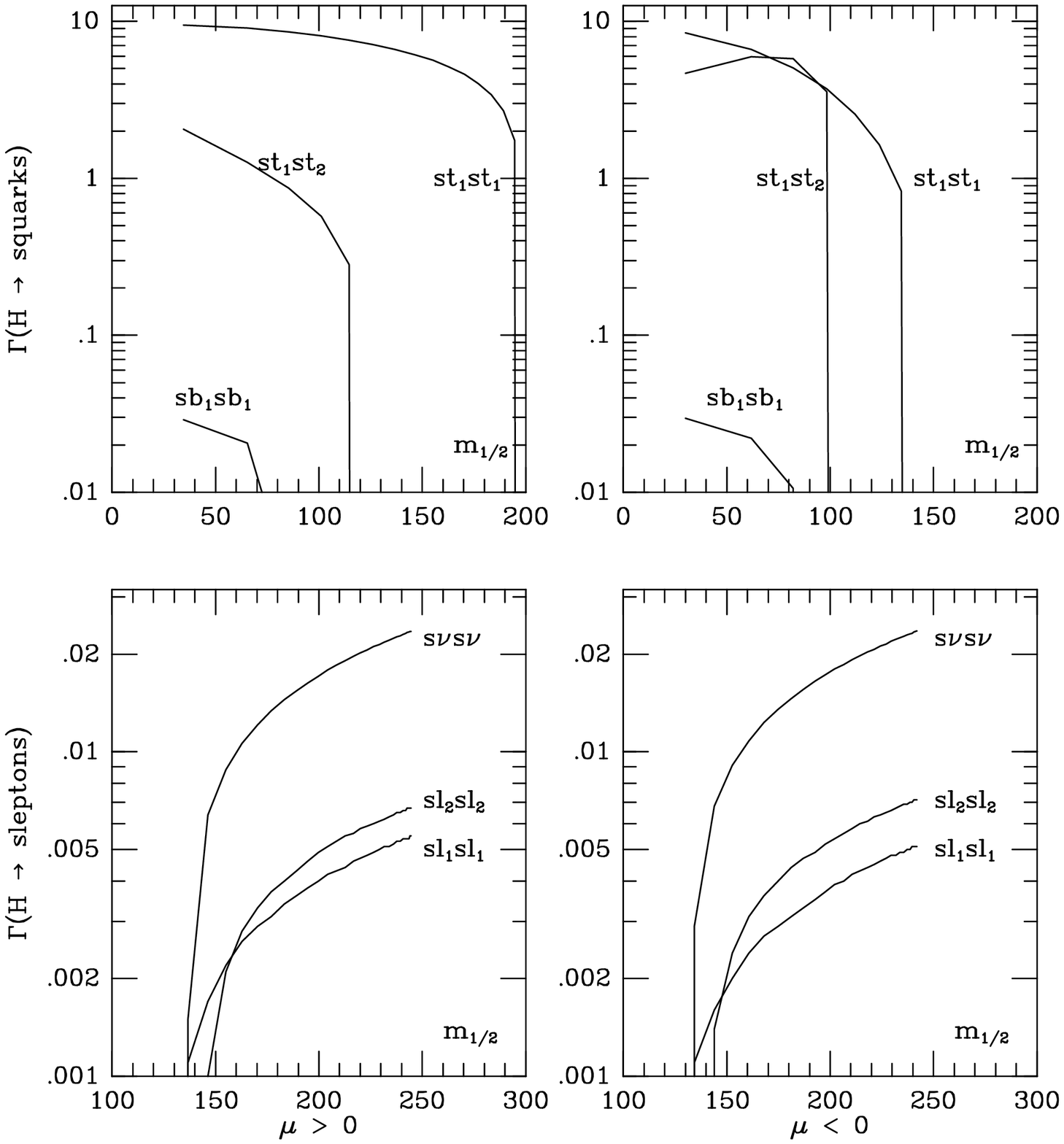,height=20.cm,width=16cm}}
\vspace*{-2.5cm}
\nn {\bf Fig.~6c:}  Partial decay widths (in GeV) of the heavy CP--even Higgs
boson $H$ into stop and sbottom squarks and into slepton pairs
as a function of $m_{1/2}$ for $\tb =1.75$,
$M_A=600$ GeV and for both signs of $\mu$. 
\end{figure}

\newpage
\begin{figure}[htbp]
\centerline{\psfig{figure=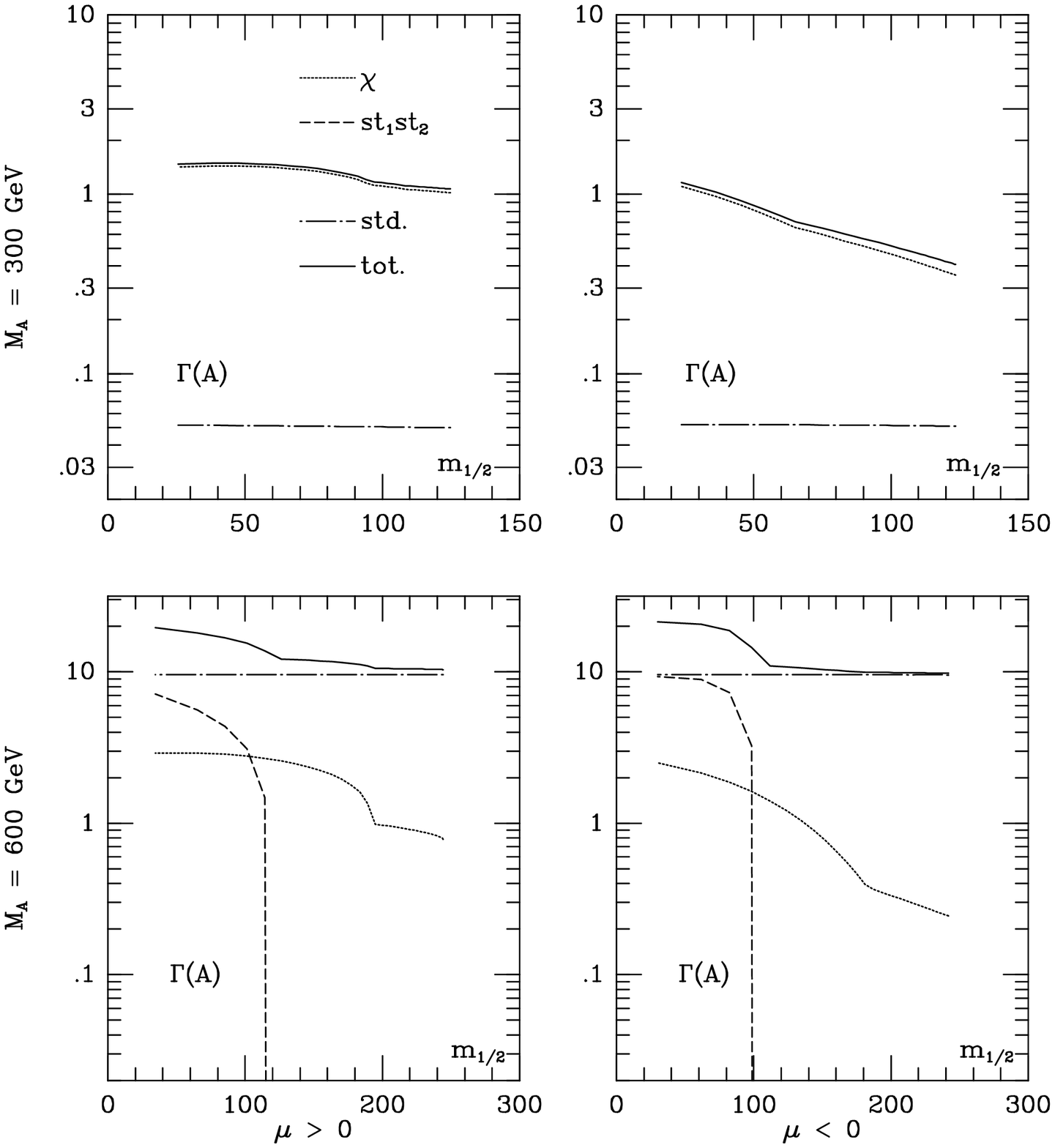,height=20.cm,width=16cm}}
\vspace*{-2.5cm}
\nn {\bf Fig.~7a:}  Decay widths (in GeV) of the pseudoscalar Higgs 
boson $A$ into charginos and neutralinos (dotted lines), stop squarks (dashed 
lines), standard particles (dott--long--dashed lines) and the total decay 
widths (solid lines) as a function of $m_{1/2}$ for $\tb =1.75$, 
$M_A=300$ and $600$ GeV and for both signs of $\mu$. 
\end{figure}

\newpage

\begin{figure}[htbp]
\centerline{\psfig{figure=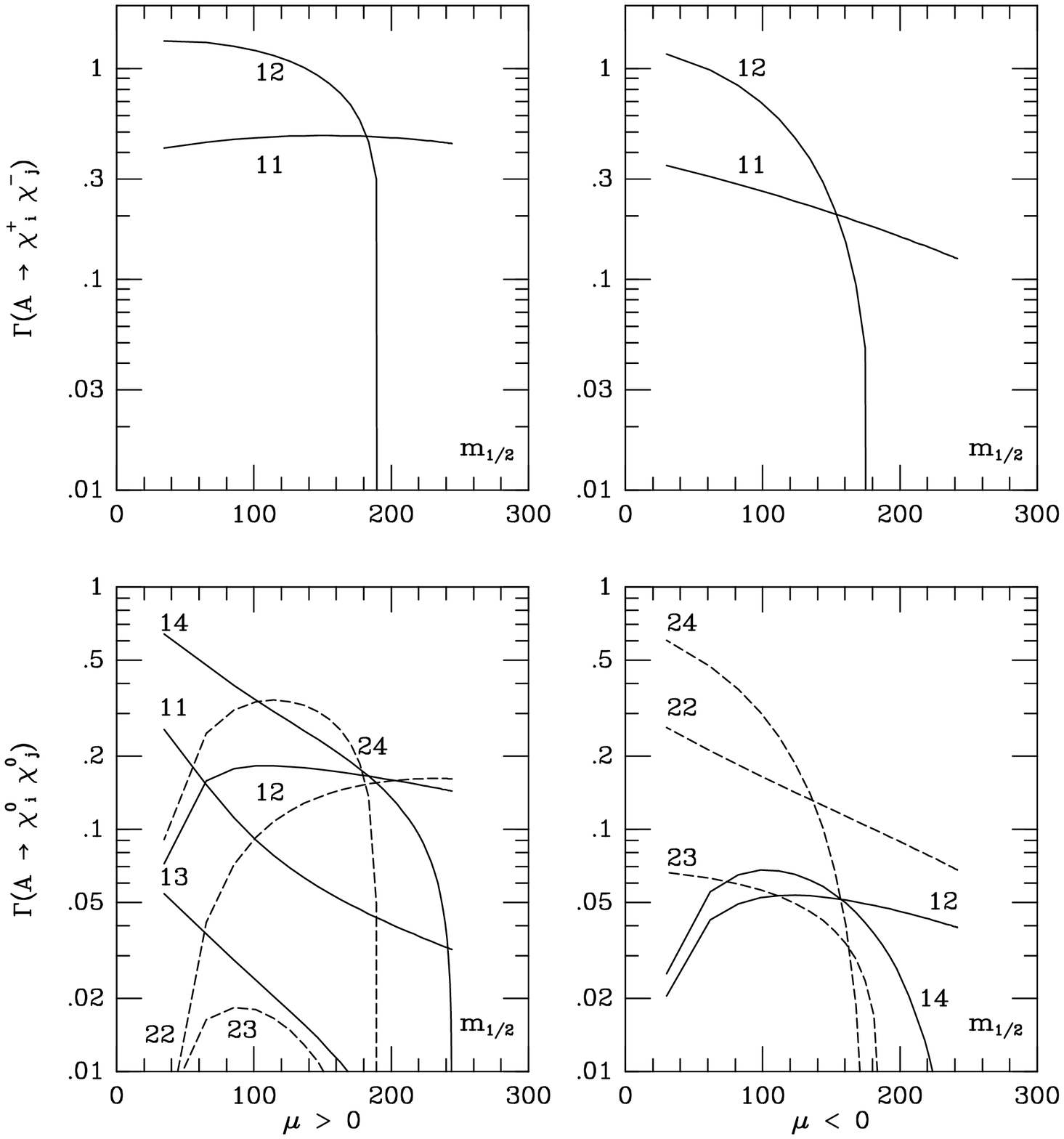,height=20.cm,width=16cm}}
\vspace*{-2.5cm}
\nn {\bf Fig.~7b:}  Partial decay widths (in GeV) of the pseudoscalar Higgs
boson $A$ into all combinations of chargino and neutralino pairs
[$ij \equiv \chi_i \chi_j$]
as a function of $m_{1/2}$ for $\tb =1.75$, $M_A=$  
$600$ GeV and for both signs of $\mu$. 
\end{figure}
\newpage 

\begin{figure}[htbp]
\centerline{\psfig{figure=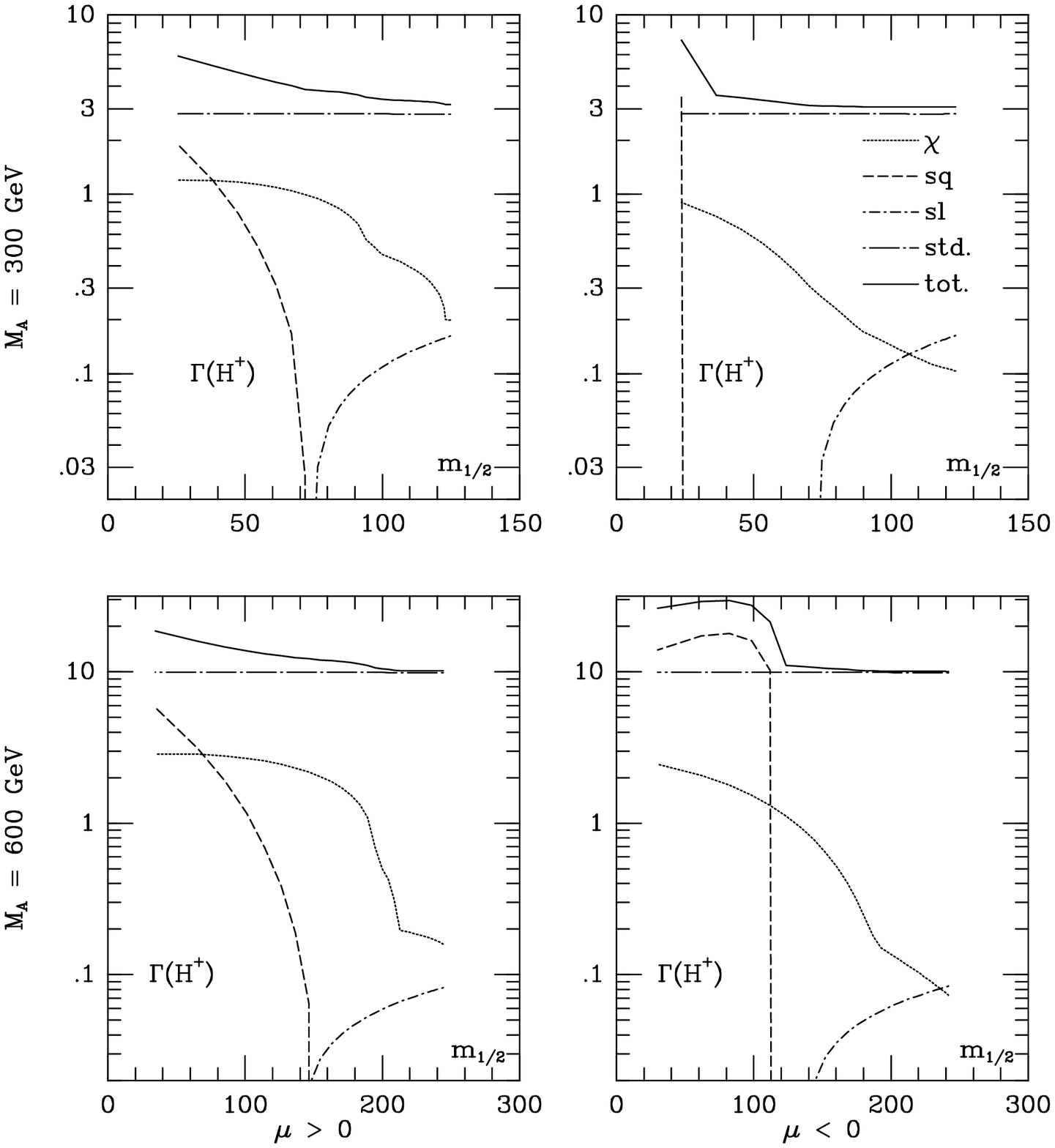,height=20.cm,width=16cm}}
\vspace*{-2.5cm}
\nn {\bf Fig.~8a:}  
Decay widths (in GeV) of the charged Higgs bosons into charginos and
neutralinos (dotted lines), squarks (dashed lines), sleptons (dash--dotted
lines), standard particles (dott--long--dashed lines) and the total decay
widths (solid lines) as a function of $m_{1/2}$ for $\tb =1.75$, 
 $M_A=300$ and $600$ GeV and for both signs of $\mu$. 
\end{figure}

\newpage

\begin{figure}[htbp]
\centerline{\psfig{figure=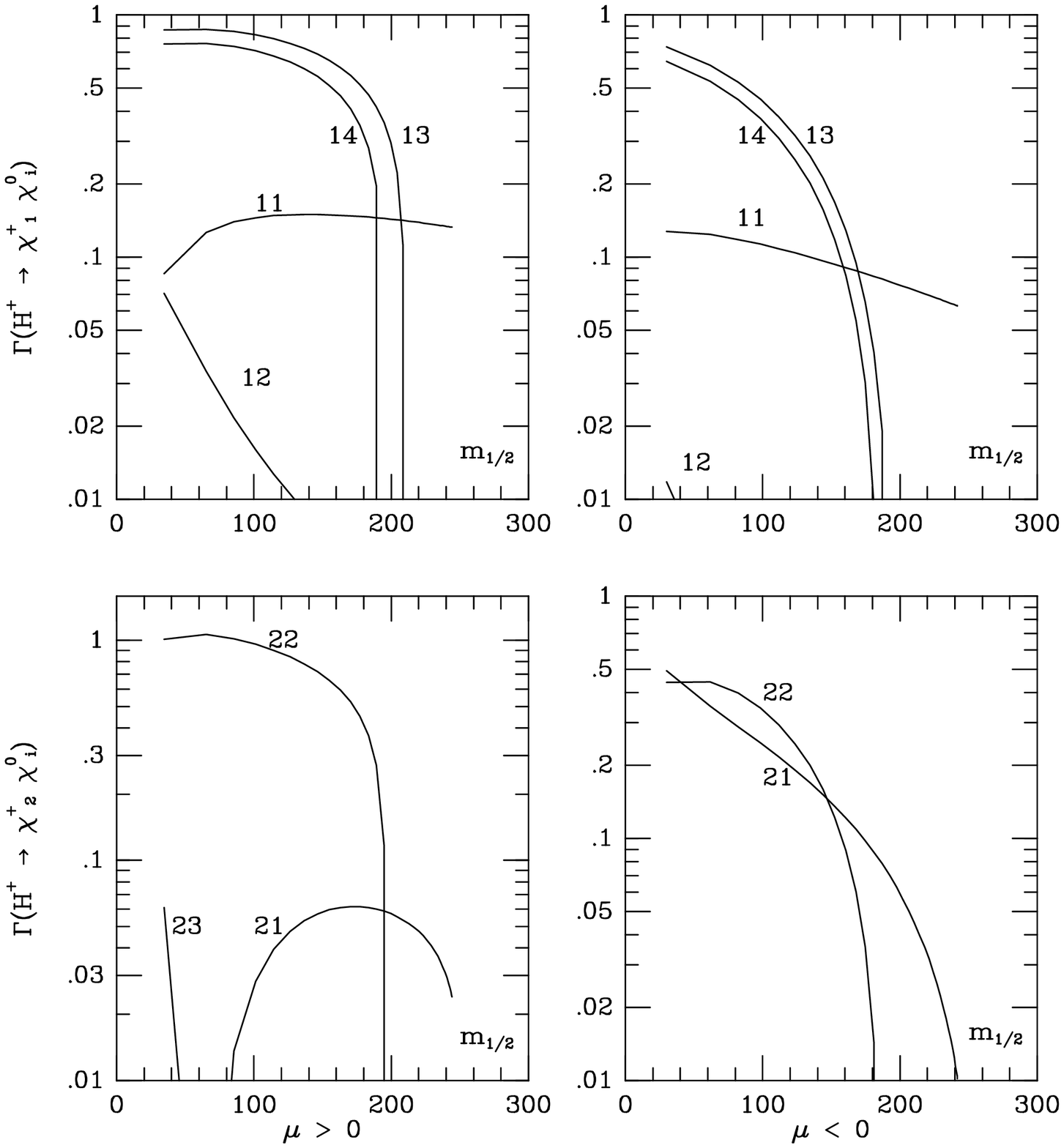,height=20.cm,width=16cm}}
\vspace*{-2.5cm}
\nn {\bf Fig.~8b:}  Partial decay widths (in GeV) of the charged Higgs 
boson $H^\pm$ into all combinations of charginos and neutralinos 
[$ij \equiv \chi_i^+ \chi_j^0$]
as a function of $m_{1/2}$ for $\tb =1.75$,  $M_A= 
600$ GeV and for both signs of $\mu$. 
\end{figure}

\end{document}